%% file: LDMX_outline.tex
\documentclass[12pt,tightenlines,superscriptaddress]{revtex4PATCH}

\usepackage[page,title]{appendix}
\usepackage{url}
\usepackage{placeins}
\usepackage{amsfonts}
\usepackage[dvipsnames]{xcolor}
\usepackage{color}
\usepackage{xspace}
\usepackage{tikz}
\usepackage[utf8]{inputenc}											
\usepackage{subfigure}

\usetikzlibrary{arrows}
\usetikzlibrary{shapes}
\usetikzlibrary{trees}
\usetikzlibrary{matrix}
\usetikzlibrary{arrows} 			
\renewcommand{\draft}[1]{{\color{blue}#1}}

\usepackage{graphicx,psfrag,bm}
\usepackage{mathrsfs,amsfonts}
\usepackage{epsfig,cancel}
\usepackage{epstopdf}
\usepackage{mciteplus}
\usepackage{latexsym}
\usepackage{amsthm}
\usepackage{amsmath}
\usepackage{amssymb}
\usepackage{hepunits}
\usepackage{hyperref}
\usepackage{bbm}
\usepackage{xfrac}
\usepackage{colordvi}
\usepackage{comment}
\usepackage{dcolumn}
\usepackage{times,wrapfig}
\usepackage{slashed}
\usepackage{booktabs}
\usepackage{relsize}
\newcommand{\be}{\begin{eqnarray}}
\newcommand{\ee}{\end{eqnarray}}

\newcommand{\benum}{\begin{enumerate}}
\newcommand{\eenum}{\end{enumerate}}
\newcommand{\bi}{\begin{itemize}}
\newcommand{\ei}{\end{itemize}}
\newcommand{\geant}{\textsc{Geant4}\xspace}

\newcommand{\ecal}{ECal\xspace}
\newcommand{\hcal}{HCal\xspace}
\newcommand{\pn}{photo-nuclear\xspace}

\newcommand{\pt}{$p_{\text{T}}$}
\newcommand{\LL}{$\langle L \rangle$}
\def\babar{\mbox{\slshape B\kern-0.1em{\smaller A}\kern-0.1emB\kern-0.1em{\smaller A\kern-0.2em R}}}

\newcommand{\ednote}[1]{} 
\newcommand{\people}[1]{} 
\newcommand{\morepeople}[1]{} 

\colorlet{RED}{red}
\colorlet{BLUE}{blue}
\colorlet{ORANGE}{orange}

\begin{document}

%
%
\title{A High Efficiency Photon Veto for the Light Dark Matter eXperiment}

\date{\today}
\hspace*{0pt}\hfill 
{\small FERMILAB-PUB-19-620-SCD, SLAC-PUB-17495}

\bigskip
\input{authorabstract}

\maketitle
\newpage
\tableofcontents
\newpage

\section{Introduction}
\label{sec:intro}
\input{sections/introduction}

\section{Signal and Backgrounds}
\label{sec:primer}
\input{sections/SigBkg}

\section{Conceptual Design of LDMX}
\label{sec:Design}
\input{sections/detectorShort}

\section{Simulation}
\label{sec:sim}
\input{sections/Simulation}
\section{Veto Methodology and Performance}
\label{sec:evtSel}
\input{sections/EvtSel}
\section{Results and Discussion}
\label{sec:res}
\input{sections/ResultsDiscussioin.tex}

\FloatBarrier
\section{Conclusion}
\label{sec:concl}
\input{sections/Conclusion}

\begin{acknowledgments}
Support for UCSB involvement in LDMX is made possible by the Joe and Pat 
Yzurdiaga endowed chair in experimental science. Use was made of the computational facilities administered by the Center for Scientific Computing at the CNSI and MRL (an NSF MRSEC; DMR-1720256) and purchased through NSF CNS-1725797. TA acknowledges support from the Royal Physiographic Society of Lund.  RP acknowledges support 
through The L’Or\'{e}al-UNESCO For Women in Science in Sweden Prize with 
support of the Young Academy of Sweden. OM, TN, PS and NT are supported 
by the U.S. Department of Energy under Contract No. 703 DE-AC02-76SF00515.
GK and NT are 
supported by the Fermi Research Alliance, LLC under Contract No. 
DE-AC02-07CH11359 with the U.S. Department of Energy, Office of Science, 
Office of High Energy Physics. BE and DH are supported by the US Department of
Energy under grant DE-SC0011925. LB acknowledges support from the Knut and Alice Wallenberg Foundation. 
\end{acknowledgments}

\clearpage
\begin{appendices}

\section{Input Variables to the \ecal BDT}
\label{sec:appEcalVar}
\input{appendix/ECalVars}

\end{appendices}

\bibliography{bibliography}

\end{document}

%% file: authorabstract.tex
\author{Torsten~Åkesson}
\affiliation{Lund University, Department of Physics, Box 118, 221 00 Lund, Sweden}

\author{Nikita Blinov}
\affiliation{Fermi National Accelerator Laboratory, Batavia, IL 60510, USA}

\author{Lene~Bryngemark}
\affiliation{Stanford University, Stanford, CA 94305, USA}

\author{Owen~Colegrove}
\affiliation{University of California at Santa Barbara, Santa Barbara, CA 93106, USA}

\author{Giulia~Collura}
\affiliation{University of California at Santa Barbara, Santa Barbara, CA 93106, USA}

\author{Craig~Dukes}
\affiliation{University of Virginia, Charlottesville, VA 22904, USA}

\author{Valentina~Dutta}
\affiliation{University of California at Santa Barbara, Santa Barbara, CA 93106, USA}

\author{Bertrand~Echenard}
\affiliation{California Institute of Technology, Pasadena, CA 91125, USA}

\author{Thomas~Eichlersmith}
\affiliation{University of Minnesota, Minneapolis, MN 55455, USA}

\author{Craig Group}
\affiliation{University of Virginia, Charlottesville, VA 22904, USA}

\author{Joshua~Hiltbrand}
\affiliation{University of Minnesota, Minneapolis, MN 55455, USA}

\author{David~G.~Hitlin}
\affiliation{California Institute of Technology, Pasadena, CA 91125, USA}

\author{Joseph~Incandela}
\affiliation{University of California at Santa Barbara, Santa Barbara, CA 93106, USA}



\author{Gordan~Krnjaic}
\affiliation{Fermi National Accelerator Laboratory, Batavia, IL 60510, USA}

\author{Juan~Lazaro}
\affiliation{University of California at Santa Barbara, Santa Barbara, CA 93106, USA}

\author{Amina~Li}
\affiliation{University of California at Santa Barbara, Santa Barbara, CA 93106, USA}

\author{Jeremiah~Mans}
\affiliation{University of Minnesota, Minneapolis, MN 55455, USA}

\author{Phillip~Masterson}
\affiliation{University of California at Santa Barbara, Santa Barbara, CA 93106, USA}


\author{Jeremy~McCormick}
\affiliation{SLAC National Accelerator Laboratory, Menlo Park, CA 94025, USA}

\author{Omar~Moreno}
\affiliation{SLAC National Accelerator Laboratory, Menlo Park, CA 94025, USA}

\author{Geoffrey~Mullier}
\affiliation{Lund University, Department of Physics, Box 118, 221 00 Lund, Sweden}

\author{Akshay~Nagar}
\affiliation{University of California at Santa Barbara, Santa Barbara, CA 93106, USA}

\author{Timothy~Nelson}
\affiliation{SLAC National Accelerator Laboratory, Menlo Park, CA 94025, USA}

\author{Gavin~Niendorf}
\affiliation{University of California at Santa Barbara, Santa Barbara, CA 93106, USA}

\author{James Oyang}
\affiliation{California Institute of Technology, Pasadena, CA 91125, USA}

\author{Reese~Petersen}
\affiliation{University of Minnesota, Minneapolis, MN 55455, USA}

\author{Ruth~Pöttgen}
\affiliation{Lund University, Department of Physics, Box 118, 221 00 Lund, Sweden}

\author{Philip~Schuster}
\affiliation{SLAC National Accelerator Laboratory, Menlo Park, CA 94025, USA}

\author{Harrison~Siegel}
\affiliation{University of California at Santa Barbara, Santa Barbara, CA 93106, USA}

\author{Natalia~Toro}
\affiliation{SLAC National Accelerator Laboratory, Menlo Park, CA 94025, USA}

\author{Nhan~Tran}
\affiliation{Fermi National Accelerator Laboratory, Batavia, IL 60510, USA}

\author{Andrew~Whitbeck}
\affiliation{Texas Tech University, Lubbock, TX 79409, USA}


\begin{abstract}
Fixed-target experiments using primary electron beams can be powerful discovery tools for light dark matter in the sub-GeV mass range. The Light Dark Matter eXperiment (LDMX) is designed to measure missing momentum in high-rate electron fixed-target reactions with beam energies of 4\,GeV to 16\,GeV. A prerequisite for achieving several important sensitivity milestones is the capability to efficiently reject backgrounds associated with few-GeV bremsstrahlung, 
while maintaining high efficiency for signal. The primary challenge arises from events with photo-nuclear reactions faking the missing-momentum property of a dark matter signal. We present a methodology developed for the LDMX detector concept that is capable of the required rejection. By employing a detailed GEANT4-based model of the detector response, we demonstrate that the sampling calorimetry proposed for LDMX can achieve better than $10^{-13}$ rejection of few-GeV photons.  This suggests that the luminosity-limited sensitivity of LDMX can be realized at 4\,GeV and higher beam energies.
\end{abstract}

%% file: sections/introduction.tex
\begin{figure}[bp]
\includegraphics[width=0.7\textwidth]{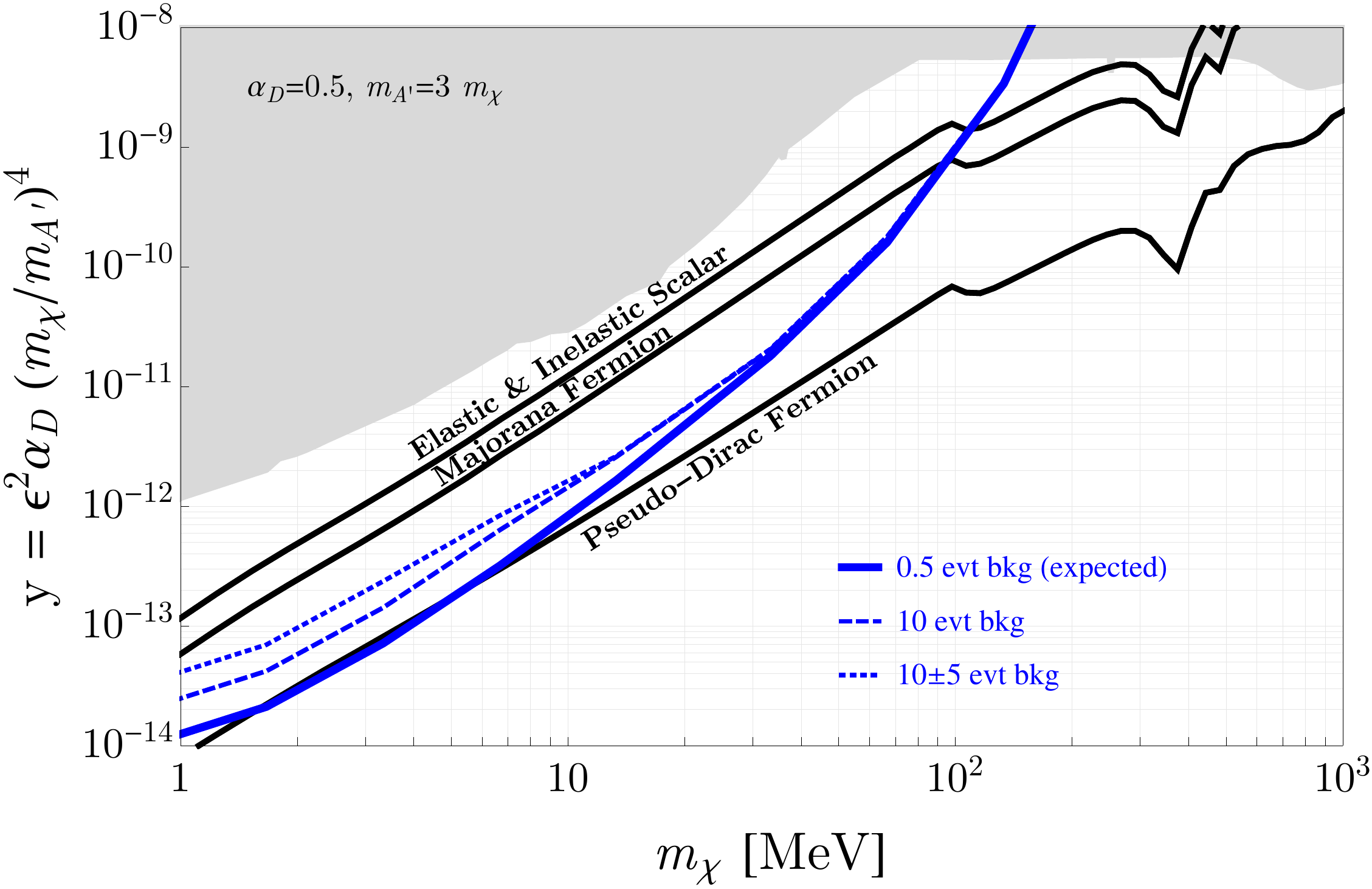}
\caption{\label{fig:reach1} 
Projected sensitivity in the $y$ vs. $m_{\chi}$ plane for an LDMX run with $4 \times 10^{14}$ electrons on target at $4 \ \GeV$ beam energy (solid blue curve), for the case of on-shell mediator production and decay into dark matter. Benchmark thermal relic targets are shown as black lines. Experimental constraints are shown for the assumption of a mediator particle mass ($m_{A'}$) three times as large as the dark matter mass and with a coupling constant $\alpha_D=0.5$ between the mediator and the dark matter. Grey regions are (model-dependent) constraints from beam dump experiments and \babar. The dashed curve shows the sensitivity in case of unexpected photon-induced  backgrounds at the 10-event level, the dotted line further assumes a 50\% uncertainty in this background. At higher DM masses, the sensitivity curves for different background assumptions converge towards the zero-background sensitivity because a transverse momentum cut efficiently reduces the background while maintaining high signal efficiency.}
\end{figure}

A compelling explanation for the origin of dark matter (DM) is that of a thermal relic from the  early universe. In this scenario, dark matter (henceforth DM) particles can have masses in the sub-MeV to 100 TeV range and must have some small non-gravitational interaction with ordinary matter.  Any such interaction implies a production mechanism for dark matter at accelerators. Scenarios where the dark matter annihilates directly to Standard Model matter (typically including electrons) are both simple and especially predictive (see \cite{Izaguirre:2014bca,Izaguirre:2015yja,Alexander:2016aln,Battaglieri:2017aum,BRNReport} for recent reviews). 
The combinations of interaction strength $y$ and dark matter particle mass $m_{\chi}$ that result in the appropriate thermal relic abundance for different types of particles are shown as black solid lines in Fig.~\ref{fig:reach1}. Probing the existence of thermal-relic dark matter in the sub-GeV mass region is well-motivated as an important part of a comprehensive search programme for dark matter. It calls for an experiment sensitive enough to explore the \emph{thermal targets} shown in Fig.~\ref{fig:reach1}, which implies interaction rates a few orders of magnitude beyond the sensitivity of current accelerator-based experiments (gray regions in Fig.~\ref{fig:reach1}).

To achieve this important goal, the ``Light Dark Matter eXperiment'' (LDMX) collaboration has developed a detector concept \cite{Akesson:2018vlm} optimized to search for dark matter particle production in high-rate fixed-target collisions of 4--16\,GeV electrons. The LDMX detector contains low-mass tracking detectors both up- and down-stream of a thin tungsten target, an electromagnetic calorimeter (\ecal{}) in the beamline downstream of the trackers, and a hadronic calorimeter (\hcal{}) surrounding the back and sides of the ECal. 

\begin{figure}[tb]
\subfigure[]{\label{fig:sigProd_direct}\includegraphics[width=0.25\textwidth]{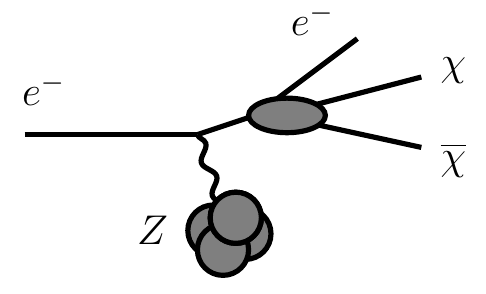}}
\hspace{0.75cm}
\subfigure[]{\label{fig:sigProd_med}\includegraphics[width=0.4\textwidth]{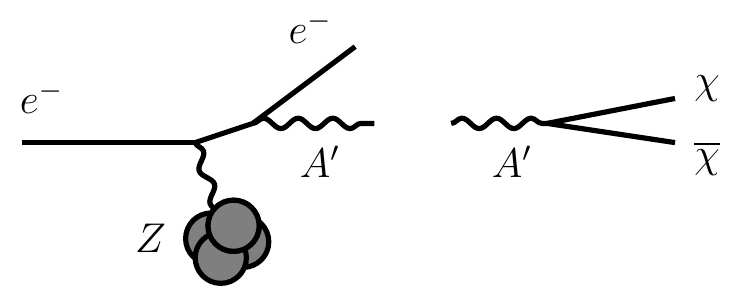}}
\hspace{0.75cm}
\subfigure[]{\label{fig:brems}\includegraphics[width=0.18\textwidth]{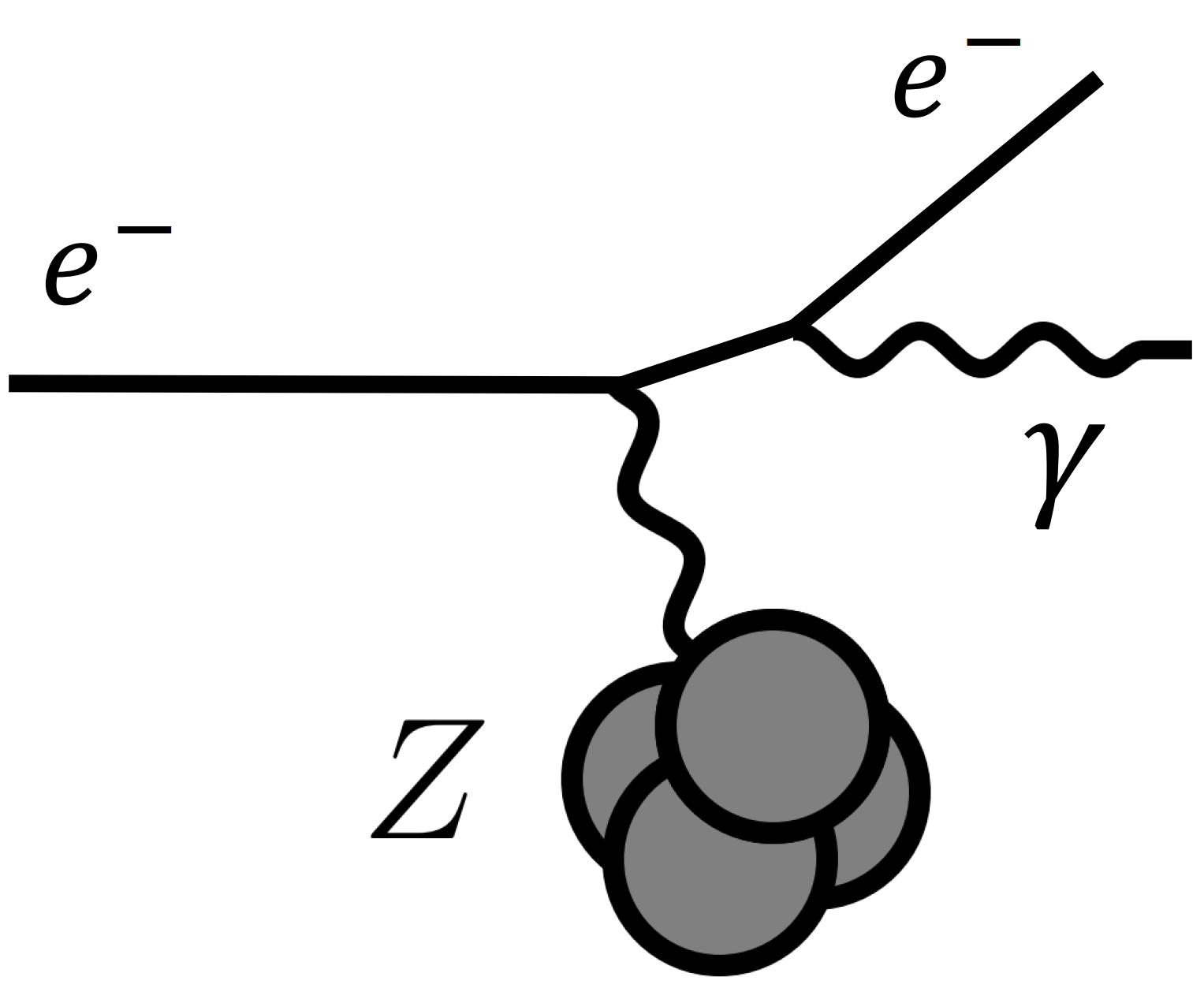}}
\caption{\label{fig:BothSignalReactions}
(a) Diagram for direct dark matter particle-antiparticle production. (b) Diagram for radiation of a mediator particle, followed by its decay into dark matter particles. (c) Standard Model bremsstrahlung of an ordinary photon. }
\end{figure}

Dark matter that couples to electrons can be produced through a ``dark bremsstrahlung'' process in the target, either directly (Fig.~\ref{fig:sigProd_direct}) or through the production and decay of a mediator particle (Fig.~\ref{fig:sigProd_med}).  The produced dark matter particles typically escape detection, leading to a clean \emph{missing momentum} signature: the electron recoils with substantially depleted energy and significant transverse momentum. The LDMX design is driven by the goal of measuring \emph{both} of these features, while efficiently vetoing background events where Standard Model particles carry the remaining energy and momentum. 
This approach is unique to LDMX and sets it apart from missing energy searches, such as NA64~\cite{Gninenko:2019qiv} at CERN, that differ from missing momentum in that they don't attempt to identify the recoiling electron or measure its transverse momentum, but instead use calorimetric measurements for a combined energy-loss measurement and background veto.

A key advantage of the missing momentum (and missing energy) approach is its high signal efficiency: for dark matter particles heavier than an MeV, more than $40\%$ of dark bremsstrahlung events pass the full signal requirements of LDMX. This stands in contrast to beam dump experiments searching for dark matter scattering in a downstream detector, which have a detection probability of $O(10^{-10})$ in the parameter space of interest. 
In this paper, we consider as a baseline an LDMX data-taking period (run) at 4\,GeV beam energy with a luminosity of $4\times10^{14}$ electrons on target (EoT). 
Thanks to the high signal efficiency, LDMX will be able to probe several important thermal targets in the range from 1 to 100\,MeV with this baseline run, if backgrounds can be kept below  $O(10)$ events. 
 
Obtaining the required performance of the background veto is the principal challenge of the missing energy and momentum approach. For a $10\%$ radiation length target, roughly 3\% of electrons will leave the target with energy low enough to be consistent with signal in LDMX.  To complete the disguise of these reactions as signal events, the other reaction products --- mainly multi-GeV bremsstrahlung photons (Fig.~\ref{fig:brems}) --- must leave little or no evidence of their passage through the LDMX calorimeters. Therefore, in the baseline $4\times10^{14}$ EoT run, approximately $10^{13}$ multi-GeV photon events must be vetoed by the detector.  These events define the \emph{photon-initiated background}. Depending on how the photon interacts, the rejection of these events relies on different veto possibilities: ordinary electromagnetic showers deposit large total energy in the \ecal; rare reactions producing muons and/or hadrons may deposit less energy in the \ecal, but with a distinctive 3D spatial profile, while also leaving easily observable hits in the \hcal and tracker. In addition to having a high signal efficiency and excellent background rejection power, the veto must be able to operate at a high event rate, because every incident beam electron leaves hits in the tracker and a shower in the \ecal.

In this paper we focus on photon-initiated backgrounds for two reasons.  First, hard bremsstrahlung in the target is the leading source of low-energy electrons in the experiment.  The photon-initiated background therefore is the largest single background for the experiment, and its rejection is crucial to LDMX's performance.  Second, photon-initiated background rejection is highly representative of the capabilities to reject background processes produced directly by the incident electron.  
Although weak interactions can produce neutrinos --- a physical source of missing energy --- the rate of signal-like weak-interaction events is negligible. 
Thus, the electron interactions with relevant rates 
are all  \emph{photon-mediated}; their rates and final states are directly related to those of corresponding photon-initiated reactions (i.e. processes from a real photon), albeit suppressed by the fine structure constant $\alpha_{EM}$.  
While beyond the scope of this paper, initial Monte Carlo studies confirm that this capability to reject photon-initiated backgrounds is sufficient to also reject backgrounds initiated by electrons at the nominal beam energy.


 Our results show that the radiation-hard fast-sampling calorimeters considered by LDMX (in combination with the trackers) can be used to efficiently reject photon-induced backgrounds from up to at least $\sim 4 \times 10^{14}$ electrons on a $10\% X_{0}$ tungsten target for a $4\,\GeV$ electron beam.  
 The most challenging event topologies are those in which a photon undergoes a rare interaction that transfers most of  
  its energy either to a single neutral hadron or to a single charged particle that decays in flight, imparting most of its energy to a neutrino and leaving only a short track in the \ecal.  
  
 These findings are also encouraging for LDMX runs at higher luminosity and beam energies, because the rates of key background topologies fall steeply with beam energy: the exclusive few-body reactions that initiate these topologies have cross-sections that fall as high powers of energy, and visible daughters from decay-in-flight are more energetic and hence more efficiently identified. Thus, for example, an $8\,\GeV$ run at 15 to 30 times greater luminosity could expect comparable background yields to the initial $4\times 10^{14}$ EoT run at 4 GeV.   Such an increase in luminosity would allow LDMX to explore much of the remaining parameter space below $\sim 1$\,GeV mass for both fermion and scalar thermal dark matter that is not accessible to other techniques. This sensitivity is independent of dark matter halo uncertainties, mediator particle mass hierarchies, dark matter spin, and other details of the interaction \cite{Battaglieri:2017aum}. 
 
 LDMX is also capable of exploring other interesting new physics, such as axion particles, millicharge particles, dark photons, and $B-L$ gauge bosons~\cite{Berlin:2018bsc}. Moreover, part of the data that LDMX plans to take will consist of electro-nuclear scattering reactions that are of direct interest to the development of physics simulations used by future neutrino experiments \cite{neutrinoPaperToAppear}.


The outline of this paper is as follows: We begin with a review of the missing momentum signature and a primer on potential backgrounds (Section~\ref{sec:primer}), followed by a description of the LDMX detector design concept (Section~\ref{sec:Design}). We then describe the simulations used in this work (Section~\ref{sec:sim}) and present our veto methodology and its performance (Section~\ref{sec:evtSel}). We conclude with a discussion of these findings and their application to the LDMX experiment (Section~\ref{sec:res}) and a brief summary (Section~\ref{sec:concl}).

%% file: sections/SigBkg.tex
\subsection{Kinematics}
%
%
%

\begin{figure}[tb]
\includegraphics[width=0.45\textwidth]{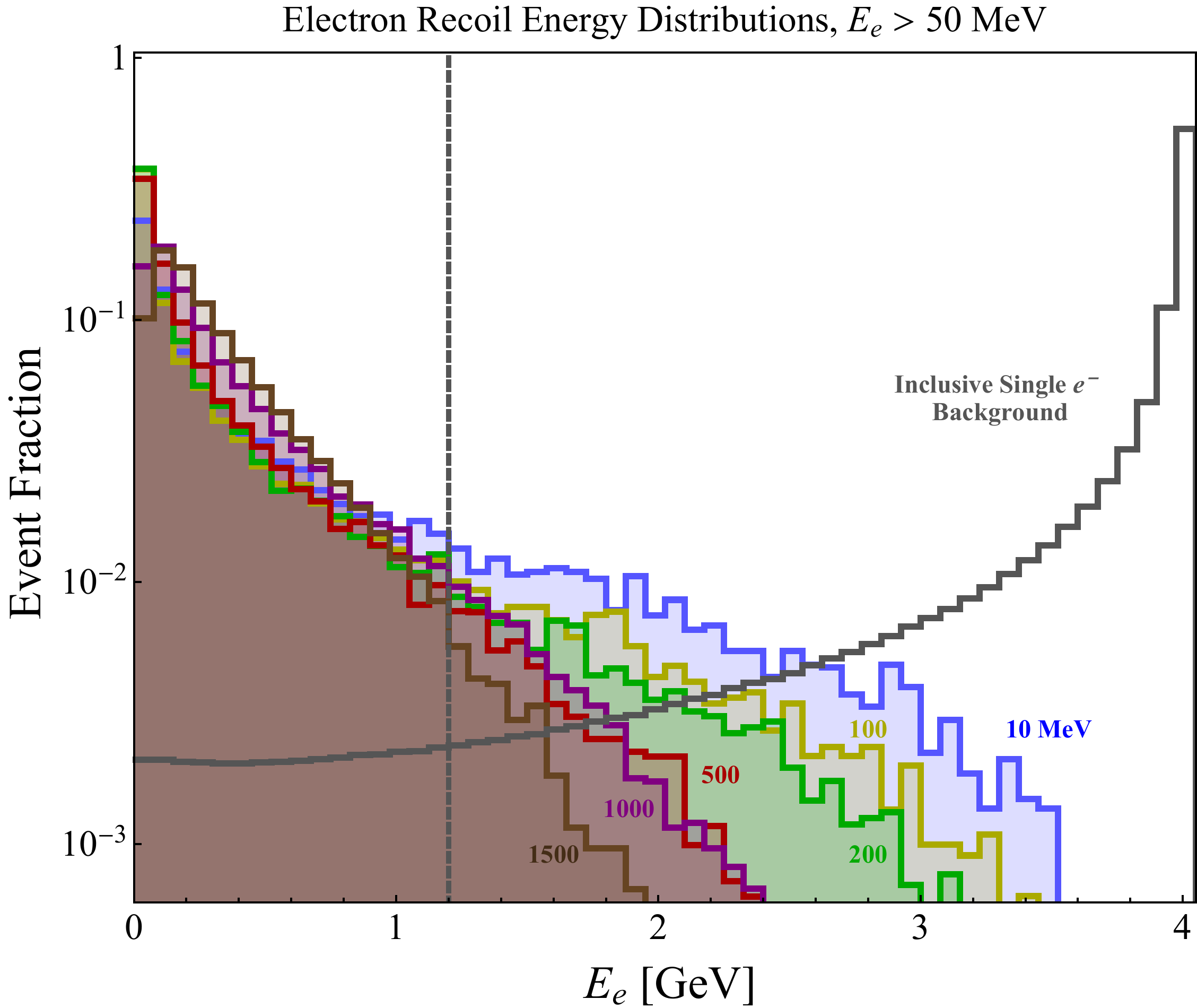}
\includegraphics[width=0.45\textwidth]{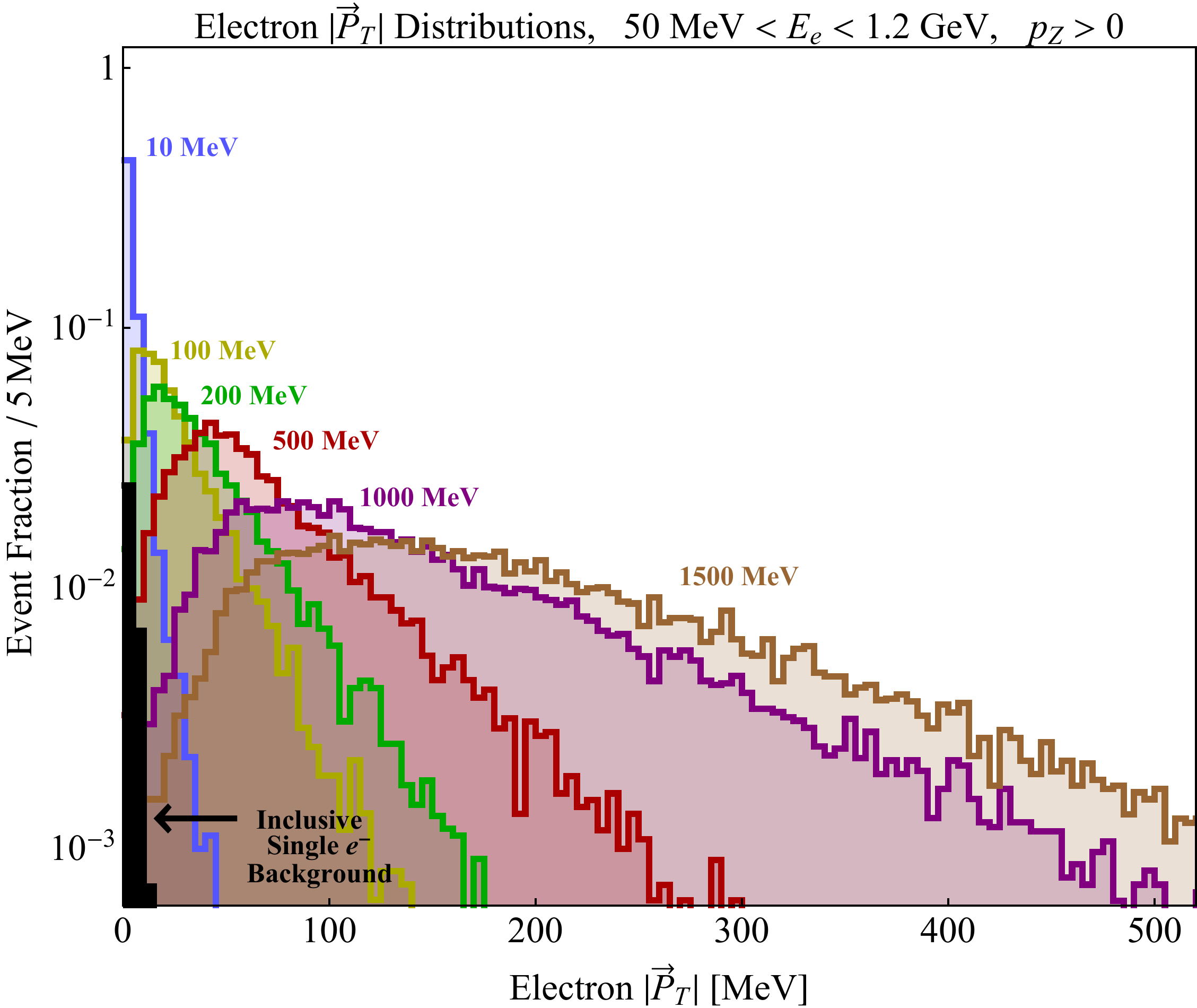}
\caption{\label{fig:LDMspectra}  
Recoil electron energy (left) and $p_T$ (right) spectra for DM pair radiation via on-shell $A'$ production (see Fig.~\ref{fig:BothSignalReactions}b) for various dark matter masses. The $p_T$ spectrum is shown after a cut on the electron energy of $50\,\text{MeV}< E_e < 1.2\,\text{GeV}$. The curves are shown for 4 \GeV incident electrons. The numbers next to each curve indicate the $A'$ mass, and each curve is re-scaled to unit area for comparison. Also included in each plot is the inclusive single electron distribution dominated by bremsstrahlung, again normalized to unit area (before selections) for comparison.}
\end{figure}

Figures \ref{fig:LDMspectra} illustrates the typical kinematic distributions of DM production reactions, and contrasts them with the kinematics of background events. For concreteness, we focus here, and in the rest of this paper, on signal reactions where DM is produced through the decay of a kinetically mixed dark photon $A'$ (see Fig.~\ref{fig:BothSignalReactions}b), following the conventions of \cite{Akesson:2018vlm}.  The distributions for DM produced through a lighter off-shell mediator or directly through a contact interaction (Fig.~\ref{fig:BothSignalReactions}a) are qualitatively similar.
The left panel of Fig.~\ref{fig:LDMspectra} shows that, for $A'$s or $\chi\bar\chi$ pairs heavier than the electron, the differential cross-section for DM production is peaked in the phase space in which the DM carries away the majority of the beam energy and the recoil electron carries relatively little~\cite{Izaguirre:2014bca}.  This behavior, which is most dramatic for higher $A'$ or $\chi\bar\chi$ pair masses, contrasts with the kinematics of SM backgrounds (shown as a gray unfilled histogram), which are dominated by photon bremsstrahlung and peak at large electron recoil energies.  

A low-energy recoil electron is the primary kinematic property for a missing energy search. For our initial study, using a 4 GeV beam energy, we will focus on events with a recoil electron  energy $<1.2\,\GeV$ (dashed vertical line in the left plot of Fig.~\ref{fig:LDMspectra}).  In addition to providing some kinematic background rejection, this selection guarantees that the additional particles produced in background reactions carry significant energy, allowing for an efficient calorimetric veto. 

The right panel of Fig.~\ref{fig:LDMspectra} shows the distribution of the electron transverse momentum (\pt{}) for events that pass this ${E_e <1.2\,\text{GeV}}$ cut for a range of DM and $A'$ masses.  These are to be contrasted with the sharply falling \pt{} distribution from bremsstrahlung, which (even after accounting for multiple scattering in a $10\%\,X_0$ target) falls off as $1/p_{\text{T}}^3$ for $p_{\text{T}} \gtrsim 4$ MeV. 
The difference in \pt{} distributions between signal and background thus offers a powerful tool, complementary to the predominantly calorimetric vetoes discussed in this paper, for mitigating background.  In the event of an excess of signal-like events, these kinematic distributions (and their correlations) would be used to assess the likelihood of various signal origins and to estimate the dark matter or dark photon mass scale.

\subsection{Detection Concept and Physics Design Drivers}
\label{sec:detconc}
\begin{figure}[tb]
\includegraphics[width=.45\textwidth]{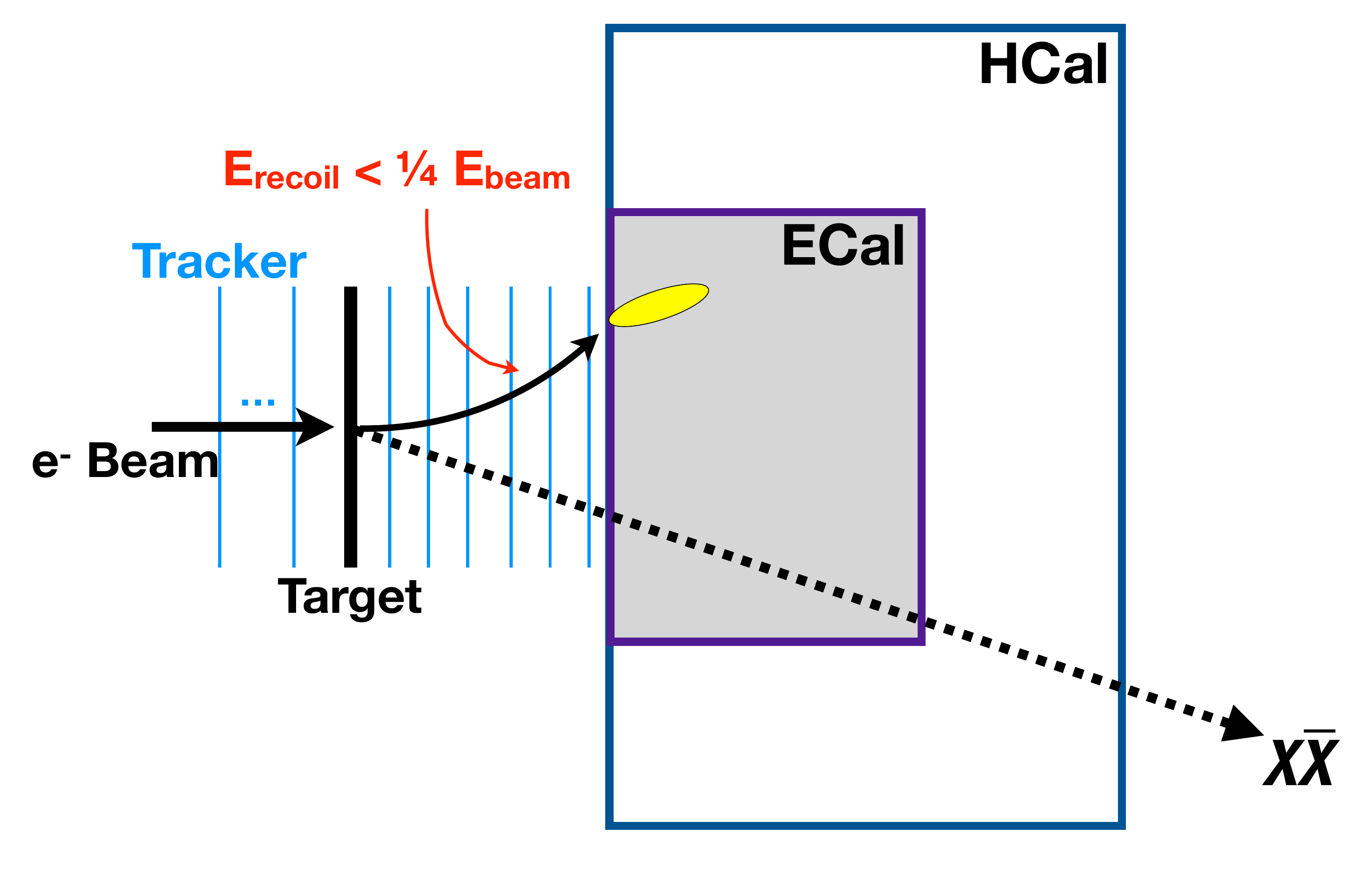}
\includegraphics[width=.47\textwidth]{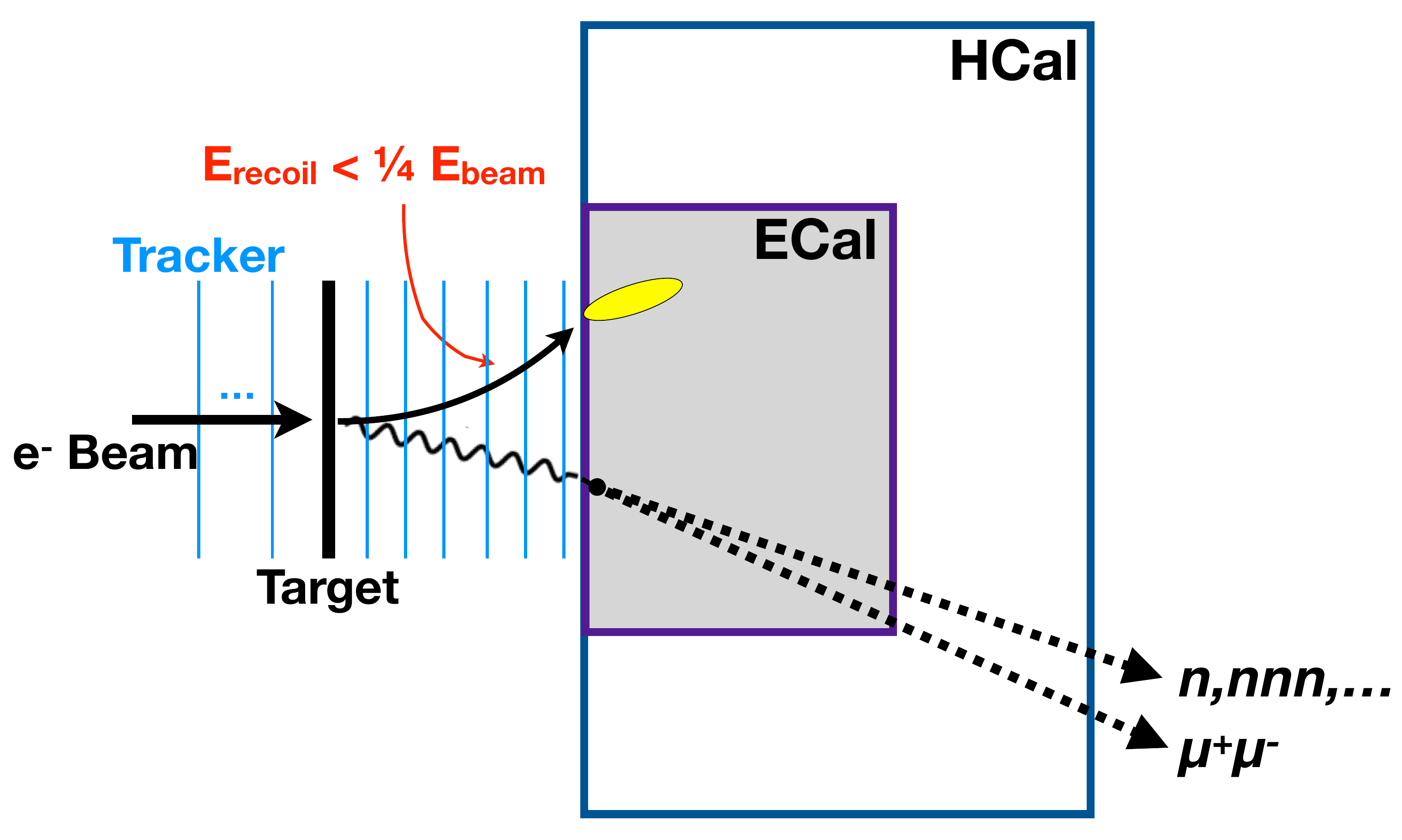}
\caption{\label{fig:detLayoutConcept} Conceptual layout of the detector with an illustration of signal (left) and an example of a hard bremsstrahlung process in the target followed by a photo-nuclear reaction in the calorimeter (right). In cases where the photo-nuclear reaction channels all or most of the total energy into neutral hadrons or other difficult to detect final states, a potential background to the signal process can arise. Developing a methodology to reject such backgrounds is the primary focus of this paper.}
\end{figure}

The detector concept shown in Fig.~\ref{fig:detLayoutConcept} is designed to exploit the aforementioned kinematic differences between signal and background, as well as enabling an efficient veto of visible particles carrying away the energy lost by the beam electron in background reactions. 
The kinematics of each beam electron are reconstructed both up- and down-stream of the target using low-mass tracking detectors in a magnetic field. The up-stream tracker tags the incoming beam electrons while the down-stream tracker measures the low-energy, moderate \pt{} recoils of the beam electrons. Calorimetry consisting of an ECal and an HCal is used to veto events with an energetic forward photon or any additional forward-recoiling charged particles or neutral hadrons. The magnetic field in the trackers, in addition to the momentum measurement, provides separation of electron and photon showers in the calorimeter.

\begin{figure}[tbh]
\includegraphics[width=0.90\textwidth]{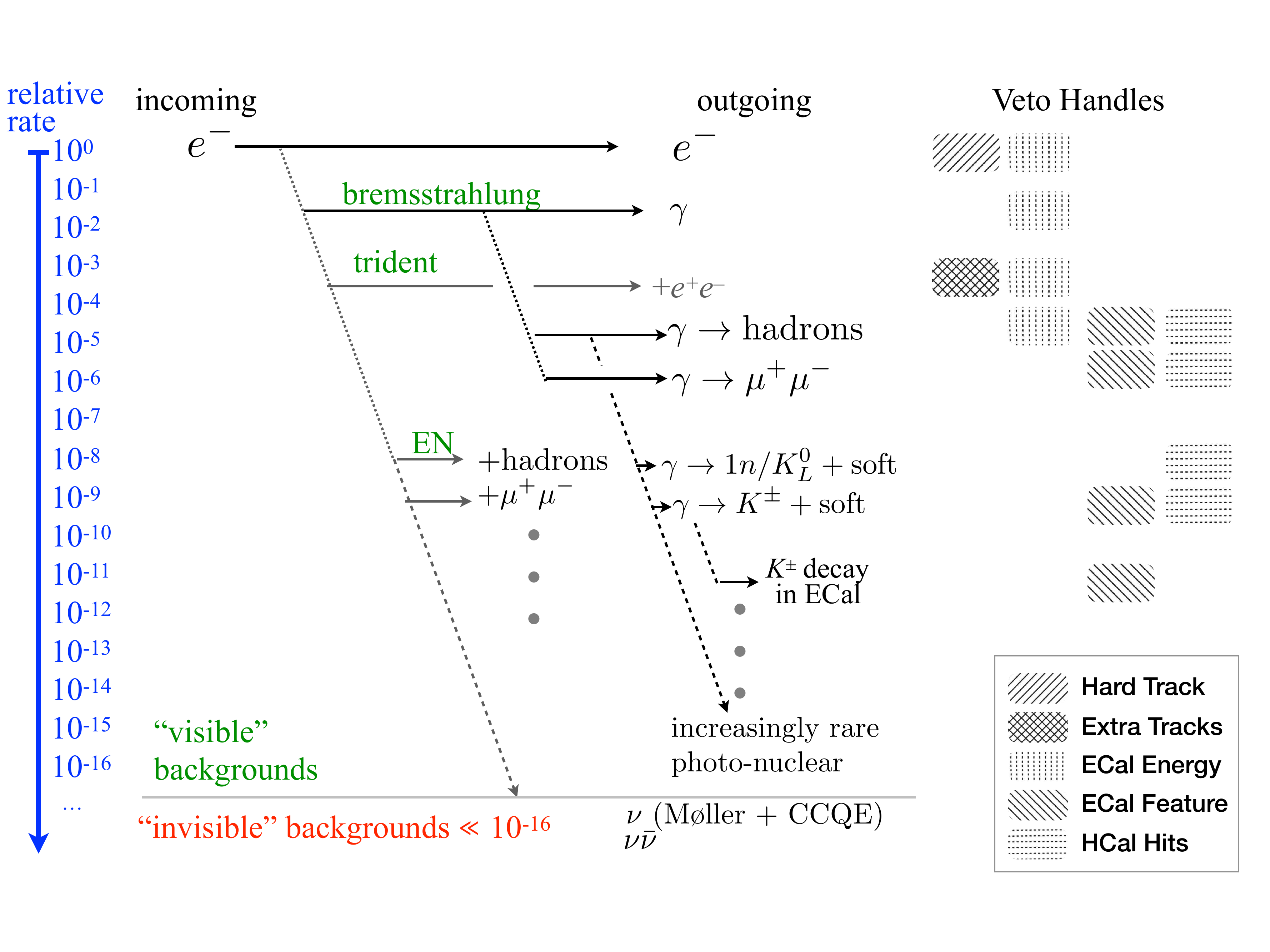}
\caption{\label{fig:BackgroundsChart} 
Flow of important (veto design driving) potential background processes and their raw rates relative to the number of beam electrons incident on the target, shown for a 4\,\GeV \,electron beam energy. On the right is a visual summary of which detector systems have significant detection and rejection power for each class of background. Not all reactions considered in our study are explicitly listed in the figure.}
\end{figure}

With this concept in mind, we can classify background events by the electron's interactions in the target, and the subsequent fate of secondary particles, as illustrated in Fig.~\ref{fig:BackgroundsChart}. 
Backgrounds from off-beam energy electrons and beam halo are efficiently rejected by the tagging tracker and spectrometer magnet, as are electrons that undergo hard interactions in the tagging tracker material. We do not consider these here, but instead focus on backgrounds arising from electrons that hit the target near the nominal trajectory, at close to the 4 GeV beam energy. 

\paragraph{\bf Non-Interacting Electrons}
Most electrons do not undergo any hard interactions in the target; they typically deposit the full beam energy in electromagnetic showers in the calorimeter. 
Because of this large energy deposition in combination with the recoil electron's high-momentum track
, such events are easily distinguished from signal.
They do, however, dominate the occupancy of and radiation dose to the detectors.

\paragraph{\bf Photon-Induced Backgrounds}
To mimic the low recoil momentum characteristic of signal events, a beam electron must interact in the target (or, with lower rate, in nearby scintillator or tracking layers).  The leading interaction of this kind is ``hard bremsstrahlung'' of a single photon above $2.8$ GeV.  
%
%
%
%
%
%
%
Typically, photons initiate a purely electromagnetic shower, depositing their entire energy in the \ecal.  
The rejection of such events relies on an electromagnetic calorimeter that is hermetic and deep enough to contain the entire shower, with fairly modest requirements on the energy resolution. 

For rare processes that interrupt the formation of an electromagnetic shower, additional techniques must be applied, which motivates the specific technology choices made for the LDMX detector systems.  
For the nominal initial beam energy of 4\,GeV, a fraction of $\sim 10^{-5}$ of the events will have a hard photon that undergoes a photo-nuclear interaction in the calorimeter, or that yields minimum-ionizing particles (MIPs), charged and/or neutral hadrons, or a combination of these, and therefore requires simultaneous \ecal, \hcal, and recoil-tracker veto capabilities. In particular, the non-standard shower profiles occurring in such events, the need to spatially resolve photo-nuclear products from the recoil electron's shower, and the production of MIPs all led to the choice of a highly granular \ecal with a high single-cell efficiency. 
Reaction types in which the hard bremsstrahlung photon transfers almost all of its energy into a single hadron pose a particular challenge, because rejection of these events relies strongly on minimizing the single-particle veto inefficiency for the produced hadron.  

These reactions typically drive the rejection required for a given particle; two such reactions of great relevance to LDMX's photon rejection capabilities are included explicitly in Figure \ref{fig:BackgroundsChart}. The rejection of events with a single, energetic neutral hadron (neutron or $K^0_L$) determines the required sensitivity and depth of the \hcal.  Events where most of the photon's energy is carried by a charged kaon, that decays in flight within the \ecal, can also be difficult to detect in the (small) region of phase space where the resulting muon is soft; from these events results the required capability of the \ecal to reject short MIP tracks, which again relies on its granularity and single-cell efficiency.

With a relative rate of $\sim10^{-6}$, the photon will convert to a muon pair.  
This final state is typically quite easily rejected, but in rare cases it also relies on the ability of the \ecal to identify short tracks, in particular when one muon is very soft and the remaining hard muon decays in-flight near the front of the \ecal. Similar considerations apply to reactions where the photon converts to pion pairs (as occurs in diffractive $\rho$ meson production), which occurs with similar rate to muon pair production.

 \paragraph{\bf Photon-Mediated and Weak Backgrounds}
Higher-order QED reactions can give rise to electron-positron pairs at the target, or a higher multiplicity of photons. These are not only suppressed relative to the single-photon reactions considered above, but also easier to reject. For example, trident reactions give rise to additional $e^+e^-$ pairs emanating from the target that can be efficiently vetoed by the tracker, \emph{in addition} to depositing an easily vetoed total energy in the \ecal.  Likewise, events with two hard photons will deposit significant energy in the \ecal unless \emph{both} photons undergo photo-nuclear reactions, which is exponentially less likely than photo-nuclear reactions of a single hard photon. 

For each rare photon interaction described above (photo-nuclear reactions and muon-conversion), there is an analogous rare electroproduction reaction mediated by a virtual photon, i.e. electronuclear scattering and muon trident production. These have a similar range of final states to the corresponding reactions of a real photon, but with a suppressed rate and a broader $p_T$ spectrum.  They also typically lead to additional charged tracks emanating from the target, which provides additional veto opportunities.  


Weak interactions of electrons are also important, as neutrino production is the only Standard Model source of prompt missing energy. However, these reactions have a very low rate at 4 GeV beam energy, with only $\sim 5$ neutrino-production events expected in $4\cdot 10^{14}$ electrons incident on a $10\%\,X_0$ Tungsten target.  These events are dominated by charged-current exchange, which has no recoil electron track and is therefore readily rejected.  Events with both a low-energy recoil track (from M\o ller scattering) accompanied by missing energy (from charged-current neutrino production) occur at a negligibly low rate, with $\sim 10^{-3}$ events produced in a $4\times 10^{14}$ EoT run.  

Combined, the above considerations motivate a detailed investigation of photon-induced backgrounds as the key to a low-background missing-momentum experiment.

%% file: sections/detectorShort.tex
\begin{figure}[!bp]
    \centering
    \subfigure[]{\label{fig:LDMX_overview}\includegraphics[width=0.42\textwidth]{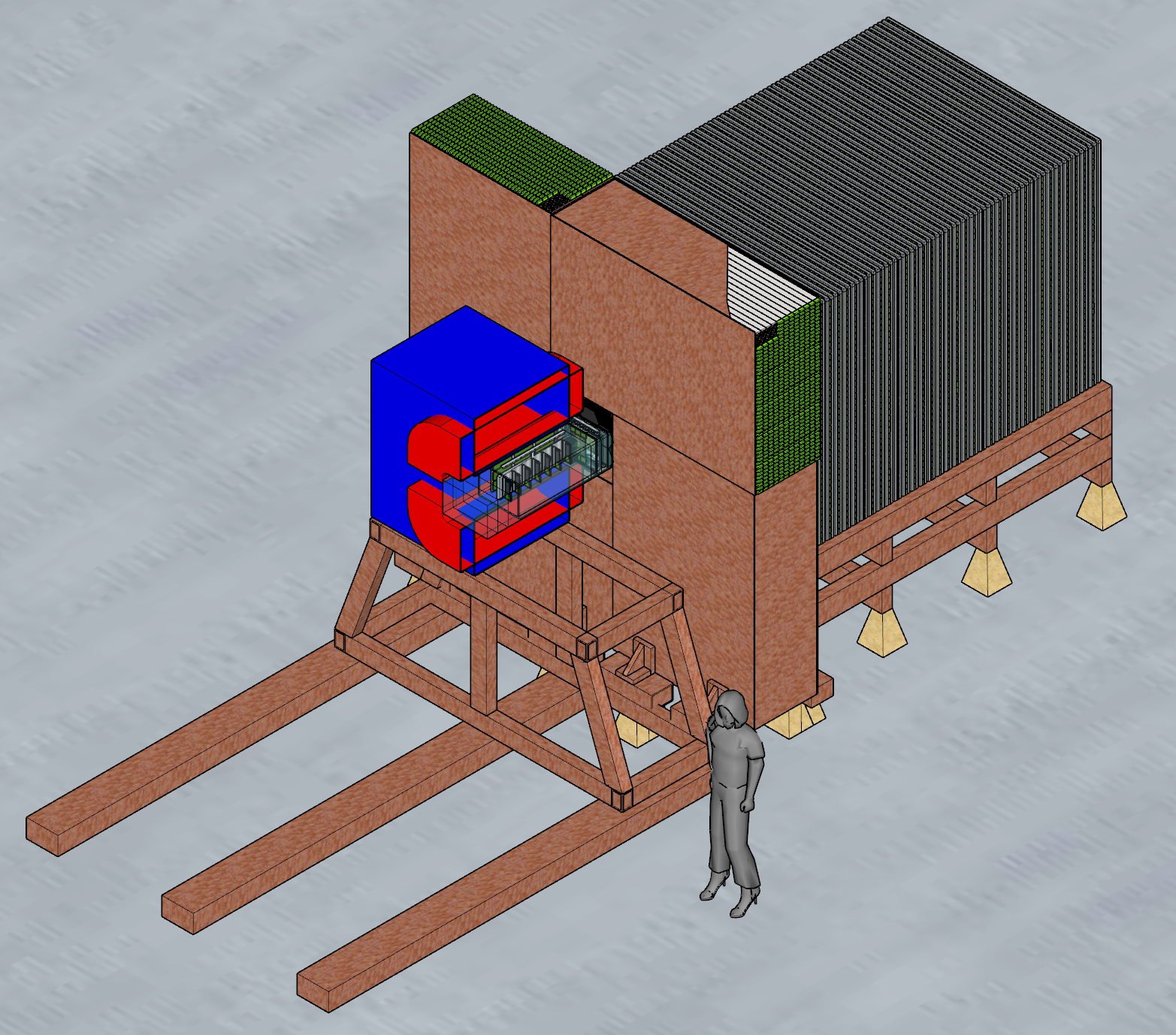}}\hspace{0.5cm}
    \subfigure[]{\label{fig:LDMX_cutaway}\includegraphics[width=0.5\textwidth]{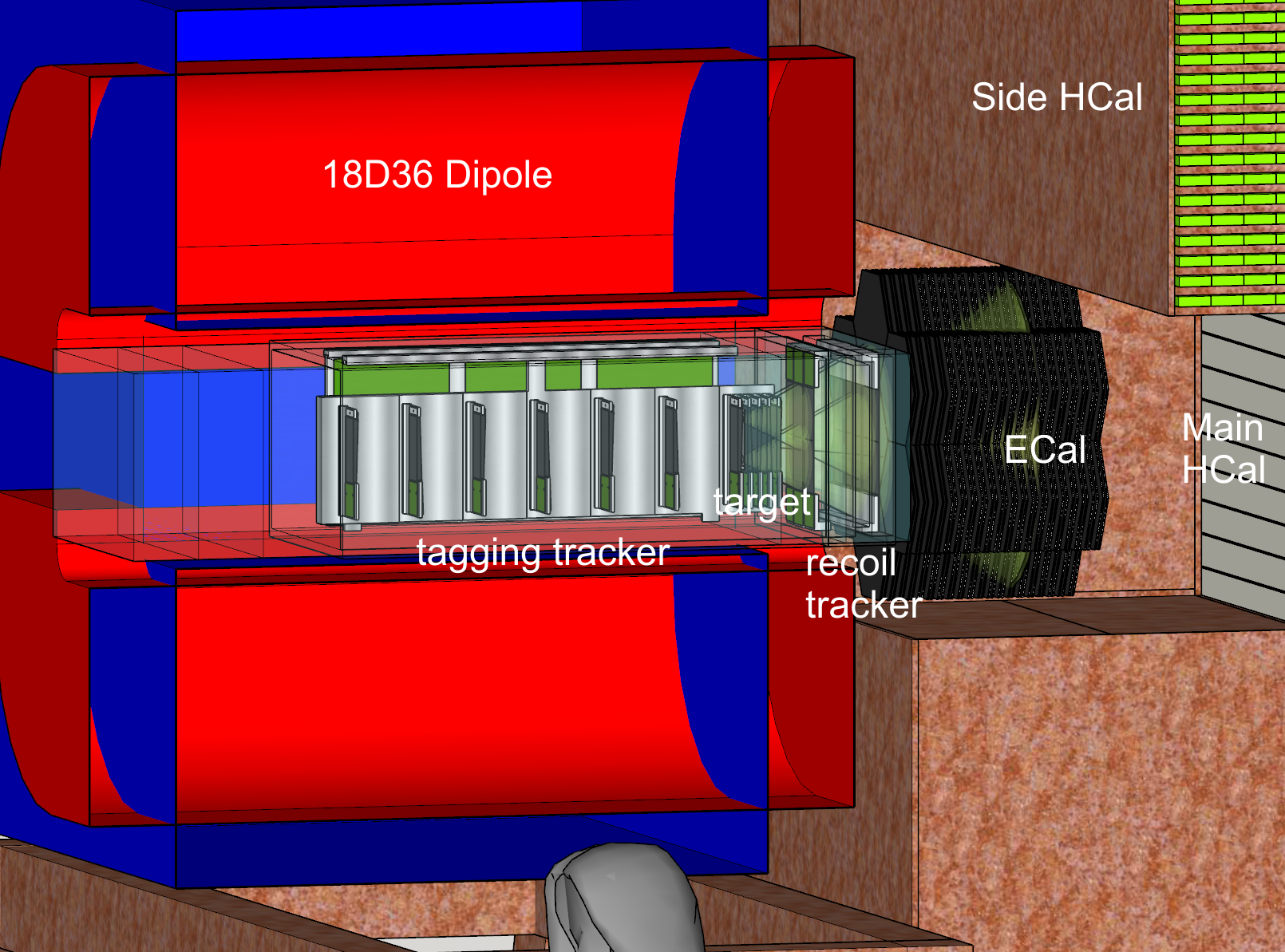}}
    \caption{\label{fig:LDMX_det}\small{(a) An overview of the LDMX detector showing the full detector apparatus with a person for scale. The tracker and \ecal are less than meter-scale, with a larger \hcal needed for neutron acceptance and containment. (b) A cutaway overview of the LDMX detector showing, from left to right, the trackers and target inside the spectrometer dipole, the forward \ecal, and the \hcal.}}
    
\end{figure}

A conceptual design for LDMX has been presented in~\cite{Akesson:2018vlm} and is illustrated in Figure~\ref{fig:LDMX_overview}. Following the beamline, its main parts are a tagging tracker inside a dipole magnet, two thin trigger scintillator planes, a tungsten target, a recoil tracker in the fringe field of the magnet, an \ecal, and behind it an \hcal.
A simulation of the design summarized below and detailed in \cite{Akesson:2018vlm} was used to produce the results presented in this paper.

\subsection{Tracking and Target}
\label{ssec:tracking}
The active elements of the tracking system are similar to those of the Silicon Vertex Tracker for the Heavy Photon Search experiment (HPS)~\cite{hps_proposal_2014}. Following the beamline, its main parts are a 60\,cm tagging tracker inside the 1.5~T field of the magnet, two thin scintillator planes, a tungsten target, and a 18\,cm deep recoil tracker in the fringe field of the magnet, as shown in Figure~\ref{fig:LDMX_cutaway}.

The target is a 350\,$\mu$m thick tungsten sheet, corresponding to 0.1 $X_0$. This thickness has not been optimized, but is expected to provide a good balance between signal rate and transverse momentum resolution from multiple scattering. 
The target is glued to a stack of two (segmented) planes of polyvinyltoluene scintillator, each 2\,mm thick, which provide a fast count of the number of incoming electrons to the trigger system.

The tracker moduless employ silicon microstrips. 
The tagging tracker consists only of stereo modules, where one sensor is oriented vertically to provide good momentum resolution, and the other is at a stereo angle of $\pm100$\,mrad. The tagging tracker places seven of these double-sided modules, spaced 10\,cm apart from each other, in a long, narrow telescope to eliminate incoming electrons with energies below the beam energy. A precise measurement of the momentum is ensured by the large spacing between the low-mass layers in the magnetic field, while the large number of layers provides the redundancy needed for high fidelity pattern recognition. 

The recoil tracker must identify and precisely measure low-momentum (50\,MeV to a few GeV) recoil electrons. To optimize the tracking for such low-momentum particles, the recoil tracker is placed in the fringe field of the magnet. The four layers closest to the target use the same stereo modules as the tagging tracker, spaced 1.5\,cm from each other, to provide good pattern recognition for very low-momentum particles. Two larger-area layers placed 9\,cm and 18\,cm downstream of the target utilize only vertically oriented strips - in single-sided modules - to provide the best momentum measurement throughout the entire momentum range.

 The efficiency of signal electron identification by the recoil tracker varies between 95\% for dark photon masses near 1\,\MeV~ to 35-50\% at 2\,GeV, as estimated from simulations.
The drop in acceptance is due to the increased polar angle and smaller momentum for high-mass signals, such that the recoil electrons are more likely to be deflected out of the recoil tracker. 

\subsection{Electromagnetic Calorimeter} 
The requirements on the LDMX \ecal are radiation hardness, fast readout, and good energy as well as spatial resolution. 
The latter is critically important for discrimination of signal from rare backgrounds that can have  extremely few and small observable energy deposits in or near the projected shower region of a bremsstrahlung photon.  These considerations lead to the selection of a high granularity silicon-tungsten (Si-W) sampling calorimeter for LDMX that makes use of the technologies developed for the High Granularity Calorimeter upgrade of the CMS experiment for the HL-LHC \cite{cms-phase-2-tdr}. 


The \ecal{}, shown in dark grey in Figure~\ref{fig:LDMX_cutaway}, is made up of 17 double layers, each containing two Si-W layers, one on either side of a cooling plane used to hold the silicon near $0^o$C. The W layers vary in thickness from $\sim$1~mm to 7~mm for a total of 14~cm, corresponding to a depth of 40~$X_0$. The 34 sensing layers have hexagonal modules arrayed in a `flower' pattern with six modules surrounding a core module.  The module differs from the CMS design in several ways. Most notably, the \ecal design will take advantage of a milder radiation environment to use sensors at least 500\,$\mu$m thick as compared to a maximum of 300 \,$\mu$m in CMS. The sensor is covered by, and wirebonded to, the CMS designed printed circuit board (`hexaboard'). The 1.5~mm thick hexaboard supports the readout ASICs. Between \ecal double layers there will be 1-1.5~mm air gaps. The ECal is a compact device fitting into a cube with $\sim$55~cm sides. 

An \ecal based on silicon pad sensors is well-suited to the identification of photons and electrons with extremely high efficiency and good energy resolution dominated by a stochastic term of 20\%/$\sqrt E$. The \ecal will use the CMS high density sensor and hexaboard designs, without modification. The module will thus have 432 individual sensing pads, each with an area of 0.56\,cm$^2$. 
The high granularity provides the ability to track charged hadrons that cross multiple silicon layers and to identify isolated charged hadrons that range out in a single layer.

The \ecal Moliere radius is found to be 27.8\,mm in simulation.  However, the lateral shower size starts out at $\sim$3~mm and only reaches the Moliere radius after traversing many sensing layers. The average layer-by-layer 68\% containment radii of EM showers, discussed in more detail below,  provide a measure of how much nearby showers must be separated to be distinguished. This is important for separation of signals due to radiated photons, recoil electrons, and beam electrons that have not interacted upstream of the \ecal.

\subsection{Hadronic Calorimeter}
The \hcal, shown in brown/green and grey in Figure~\ref{fig:LDMX_det}, is a scintillator-based sampling calorimeter comprising a large number of nuclear interaction lengths of absorber. Its main function is to detect neutral hadrons - mostly neutrons produced in the target or \ecal{} - in the energy range from approximately $100~\MeV$ to several $\GeV$ with very high efficiency, e.g. for neutrons with an energy of a few GeV, the veto inefficiency should not exceed $10^{-6}$. The \hcal{} must also measure the component of electromagnetic showers escaping the \ecal, and be sensitive to MIPs, such as muons. 

The \hcal dimensions and segmentation are being optimized with detailed GEANT simulations, including several background processes and single particle response. The \hcal design used for the results presented here consists of a steel-scintillator calorimeter of approximately 90 layers of 25 mm absorber plates and 15 mm thick scintillator, totaling 15 nuclear interaction lengths or 4\,m. The scintillator is doped polystyrene, segmented into 3\,m long bars that are 15~mm thick $\times$ 50~mm wide and are co-extruded with an integrated TiO$_2$ reflector. The bars have a through-hole into which a wavelength-shifting fiber is inserted, which is read out at the ends with Silicon-Photomultipliers (SiPMs). This design is similar to what is used in the Mu2e Cosmic Ray Veto system~\cite{Atanov:2018ich,Artikov:2017lsc}. The bars are arranged in an ($x,y$) configuration, with the longitudinal position within a bar constrained by means of timing correlations for signals arriving at the two ends of each bar. The transverse dimension of the \hcal{} therefore corresponds to the 3\,m length of the bars. As the \hcal rates are low, this design does not suffer from significant occupancy issues. 

Reaction products at large angles are detected by a smaller device surrounding the \ecal, termed the Side \hcal, with a similar design but finer sampling and single-ended readout due to space constraints. 



%% file: sections/Simulation.tex
\subsection{Software Framework}
Signal events are generated with a specialized version of MadGraph/MadEvent (MG/ME)~\cite{alwall:2007st} that includes nuclear form factors, while background processes are produced directly in 
\geant~\cite{Agostinelli:2002hh,Allison:2016lfl}, with modeling improvements and validations described below. The propagation of particles
through the detector along with their interactions with material is simulated
using an object-oriented framework,  
which wraps a custom version of the \geant toolkit ({\tt 10.02.p03}) and adds several services as well as an analysis/reconstruction pipeline. 
The latter includes stages that simulate the detector electronics for all event model hits, adds noise, performs basic track finding and can be used to apply various vetoes optimized to reject backgrounds. 
 
Event generation is based mainly on sources provided by \geant using the physics list {\tt FTFP\_BERT} augmented by the list {\tt G4GammaConversionToMuons}. 
The {\tt FTFP\_BERT} list contains all standard electromagnetic processes and makes use of the Bertini cascade~\cite{BertiniCascade} to model hadronic interactions. Although the Bertini cascade model has been validated up to incident energies of 9 GeV (see e.g. \cite{Banerjee:2011zzf}), by default the {\tt FTFP\_BERT} physics list transitions between the Bertini cascade and the Fritiof string model~\cite{Andersson:1986gw} at 3.5 GeV energy. Because the latter provides a poor model of the intranuclear physics that produces soft hadrons, and to avoid an unphysical discontinuity in the energy range of interest for LDMX, the physics list was modified to use the Bertini model throughout the LDMX phase space. Specific validations of and improvements to the Bertini photo-nuclear model are described below. 

The simulation of the \ecal uses \geant  coupled with an electronics simulation that converts the energy deposits in the silicon layers into simulated digital hits. Noise is introduced at the digitization stage using a realistic model based on tests performed with prototypes of the CMS front-end readout chip.

The \hcal simulation uses \geant combined with a model of the scintillator and readout responses to convert the energy deposited inside the active medium into photo-electrons (PEs). The scintillator response is simulated as a  Poisson-distributed number of PEs based on the energy deposited in each scintillating bar. A conservative estimate of ${15\,\text{PE}/\text{MeV}}$ of energy deposited in the bar was used; future simulations will be based on the photo-electron yield measured by the Mu2e experiment and an updated bar geometry. These values include quenching effects on the scintillation light yields (Birks' law), parametrized from earlier measurements of plastic scintillators~\cite{NIM80.2.239}.  Uniform Poisson-distributed noise is added to the \hcal signals.

\subsection{Simulated Event Samples}
\label{sec:simSamples}
The simulation of signal events is based on an $A'$ model that has been used 
for the APEX test run~\cite{Abrahamyan:2011gv} and the HPS 
experiment~\cite{PhysRevD.98.091101}, and to which light DM particles are added.
While this paper
focuses on the on-shell production process, the kinematics for DM particle 
production via an off-shell $A'$ would be very similar. The simulation of signal
events proceeds in two steps: First, the DM production process
$e^-W\rightarrow e^-WA' (A'\rightarrow \chi\bar{\chi})$ is simulated within
MG/ME, where $W$ represents the tungsten nucleus 
(assumed to be at rest initially) and $\chi$ the DM particle. The generated
events are written to LHE files that are then passed to \geant in the second 
step to simulate the detector response. Samples of $1.5-3\times 10^{6}$ events were produced for  $A'$ masses of 1, 10, 100, and 1000 MeV spanning the mass range of interest for LDMX.


The samples used for the study of background processes were generated directly 
in \geant, with  modifications to achieve better accuracy of the modeling 
of certain final states, as described below. 
A particular focus lies on few-body background processes that are limiting for LDMX.  The rates and kinematics of these few-body reactions on low-Z nuclei are well-measured~\cite{Mirazita,Anghinolfi:1985bi,
Pomerantz:2009gf,Bartholomy:2004uz,Zhu:2004dy,Anderson:1969bq,Alvarez:1968zza,
Alanakian:1979gd,PhysRevC.20.764,Degtyarenko:1994tt,StepanPrivateCommunication,Gudima:2006pn,McCracken:2009jp}, allowing a reliable estimate of the required veto efficiency.  

The samples considered in this study are one nearly-inclusive sample of $1.3\times 10^9$ EoT 
and four high-statistics exclusive samples of event topologies where a hard photon radiated in the target undergoes a specified rare interaction in either the target or the \ecal. 
A combination of bremsstrahlung preselection (filtering) and the \geant occurrence biasing toolkit is used to efficiently generate the latter samples at equivalent statistics of $2 \times 10^{14} - 2\times 10^{15}$ EoT, comparable to the $4\times 10^{14}$ EoT baseline luminosity used throughout this paper. All simulations assume a beam energy of 4\,GeV. 

To study the energy deposited by background events in the \ecal, we use the nearly-inclusive sample, with the electrons fired from upstream of the target.  Electromagnetic (EM) and PN processes are included in the \geant physics list at their physical rates, while muon conversion and electronuclear (EN) reactions are disabled. Because this sample is used for \ecal-specific studies, a small fraction of events where energetic secondaries miss the \ecal in a manner that can be efficiently vetoed by other detectors are excluded from the final sample. 

The other four samples are used to study in detail the detector response to rare photon-induced backgrounds.  For these samples, a hard-bremsstrahlung filter restricts the
simulation to events where the electron undergoes photon bremsstrahlung in the target and 
leaves the target with an energy of less than 1.5\,GeV. Further selections on 
the fate of the photon are then applied, allowing for the generation of 
exclusive samples of photons undergoing either PN reactions or muon conversion, in either the target or the \ecal.
The latter reactions have cross-sections at the level of $10^{-3}-10^{-5}$ of the conversion cross section.  To simulate these rare reactions efficiently, we again use the occurrence biasing toolkit introduced in \geant version 10.

Because we use \geant as an event generator for rare background reactions, we have carefully validated and improved several aspects of the modeling.  The \geant Bertini Cascade model is based on a cascade of individual particle-nucleon (and particle-dinucleon) interactions, for which total cross-sections and 2-body final state kinematics are taken from data. The accuracy of the Bertini cascade model in modeling multi-GeV hadron-nucleon collisions was validated by the \geant team in e.g.~\cite{Banerjee:2011zzf}.
The inclusive PN samples serve two roles in LDMX studies: first, to validate the expectation that single-particle-like reactions (where most of the photon's energy is carried by a single hadron) are the most challenging to veto, and second to sample the phase space of these reactions.  To ensure that the latter goal is accurately achieved, we have validated the rates and phase-space distribution of event types that are design drivers for LDMX, both at the level of inputs to the Bertini cascade model and output events.  These reactions include single forward neutrons and neutral kaons (which drive the depth needed for the HCal), moderate-angle neutron pairs (which drive the width needed for the back HCal), and single forward charged kaons and pions (which drive the depth segmentation and MIP rejection capabilities of the \ecal).  Compared to  data (including \cite{Mirazita,Anghinolfi:1985bi,
Pomerantz:2009gf,Bartholomy:2004uz,Zhu:2004dy,Anderson:1969bq,Alvarez:1968zza,
Alanakian:1979gd,PhysRevC.20.764,Degtyarenko:1994tt,StepanPrivateCommunication,Gudima:2006pn,McCracken:2009jp}), the single- and di-neutron reactions, as well as backscattered hadrons in the so-called “cumulative” region, were found to be significantly overpopulated by the default Bertini model; by contrast, single-pion and single-kaon final states were 
underpopulated. The origin of each discrepancy was identified and corrections implemented, many of which have also been incorporated in subsequent official \geant versions. 

Similarly, our validation revealed that the \geant model of photon conversions to $\mu^+\mu^-$ via coherent scattering off a nucleus used an unphysical nuclear form factor. 
In addition, approximations to the phase-space distribution of outgoing muons
were found to be inaccurate on the tails. The dimuon samples used in this study
were therefore generated with a modified version of the {\tt G4PhotonConversionToMuons} class, which uses the full dimuon phase space distribution from eqn. (2.3) of Ref.~\cite{Tsai:1974}, assuming elastic recoil of the nucleus and keeping only the $W_2$ component of the result. As a test of the implementation, the muon kinematics from the resulting simulation was validated against a MadGraph/MadEvent event sample (with the same approximations in modeling of the nuclear couplings), and found to be in excellent agreement even on the tails of the distribution.

The samples used in this analysis and the corresponding statistics are listed in 
Table~\ref{tab:bkg_samples}. 

\begin{table}[htbp]
    \begin{tabular}{ r | c | c } 
        \hline
        \hline
        \textbf{Simulated sample} & \textbf{Total events simulated} & \textbf{EoT equivalent} \\
        \hline
        Inclusive EM + PN              & $1.3\times 10^{9}$   & $1.3\times 10^{9}$ \\
        Target $\gamma \to \mu \mu$    & $6.3\times10^{8}$    & $8.2\times 10^{14}$ \\
        \ecal $\gamma \to \mu \mu$     & $8\times10^{10}$     & $2.4 \times 10^{15}$ \\
        Target \pn                     & $8.8 \times 10^{11}$ & $4.0 \times 10^{14}$ \\
        \ecal \pn                      & $4.7 \times 10^{11}$ & $2.1 \times 10^{14}$ \\
        \hline
        \hline
    \end{tabular}
    \caption{The dedicated simulated background samples used in this analysis along with the
             corresponding statistics.  The listed detector volume specifies the origin where 
             the interaction is simulated to occur. The total events simulated corresponds to
             the total electrons fired on target in the simulation.  The biasing factor passed
             to the \geant occurrence biasing toolkit is used to scale the total events simulated
             to the electron on target (EoT) equivalent.} 
    \label{tab:bkg_samples}
\end{table}

%% file: sections/EvtSel.tex
This section presents a series of event selection criteria designed to reject all photon-initiated backgrounds, regardless of where and how the photon interacts. As noted in Sections \ref{sec:intro} and \ref{sec:detconc}, this veto is also expected to efficiently reject photon-mediated background reactions, which have closely related final states. We present the cutflow in a sequence motivated by the background flow chart (Fig.~\ref{fig:BackgroundsChart}), with each successive cut efficiently rejecting increasingly rare background reactions. 

Sec.~\ref{ssec:trigger} presents the \ecal missing-energy requirements and primary physics trigger.  By studying the reconstructed energy distribution in the inclusive event sample, we illustrate that the non-Gaussian missing-energy tail is dominated by events involving a rare photon reaction.  

Requiring exactly one track with $p<1.2 \,\GeV$ (Sec.~\ref{ssec:tracksel}) selects events where the beam electron undergoes a hard emission in the target.  The sample of events surviving the trigger, missing energy, and track momentum cuts is dominated by events containing a hard bremsstrahlung in the target, followed by a rare reaction of the primary photon.  
The majority of these events can be rejected based on detailed characteristics of the energy deposition in the \ecal, such as transverse and longitudinal shower shapes and the energy deposited in isolated hits.  These features, which take advantage of the excellent granularity and single-hit efficiency of the \ecal, are combined into a boosted decision tree (BDT) to maximize signal-to-background discrimination.  Sec.~\ref{ssec:bdt} (see also Appendix \ref{sec:appEcalVar}) summarizes the \ecal features and the BDT performance, focusing on the background sample that dominates after the preceding selections --- events where the bremsstrahlung photon undergoes a photo-nuclear reaction in the \ecal.   

A veto on energy deposition in the \hcal (Sec.~\ref{ssec:hcalveto}) 
suppresses events where the reaction products emerge from the target at wide angles, outside the \ecal acceptance, or those where most of the incident photon's energy is carried by a hard neutron or $K^0$ that does not interact in the \ecal.  

One class of events survives this sequence of vetoes: those where most of a photon's energy is carried away from a photo-nuclear reaction by a single charged kaon, which in turn decays in-flight giving most of its energy to a neutrino.  Rejection of these reactions requires the identification of short tracks in the ECal, discussed in Sec.~\ref{ssec:trackveto}.  

\subsection{ECal Missing Energy: Trigger and Offline Selection} \label{ssec:trigger}
Signal events with a low-energy electron and no other visible particles are expected to deposit significantly less energy in the \ecal than typical background events, in which the full energy of the beam electron goes into electromagnetic showers in the \ecal.  This motivates a physics trigger based on online reconstruction of the \ecal energy.   
The primary trigger sums the energy depositions from the first 20 layers, where the majority of the energy is deposited for typically-developing showers. This choice provides sufficient rate reduction and allows for unbiased sideband studies based on the energy in the late layers.  The trigger sum energy is combined with information about the number of electrons in a bucket, as determined by counting hits in the near-target scintillator array.  

In single-electron events, the online \ecal energy is required to be $<1.5$ GeV to trigger, corresponding to at least 2.5\,GeV of missing energy.  This threshold maintains acceptance for signal events with upward fluctuations in the reconstructed energy of the recoil electron, while adequately suppressing the trigger rate.  Missing-energy thresholds of 2.3--2.7 GeV are envisioned for 2- and 3-electron events~\cite{Akesson:2018vlm}.  In the studies presented here, we apply the single-electron trigger requirement uniformly without including the effects of pile-up. 
Similar logic underlies the offline \ecal selection --- based on the energy deposited in the full depth of the \ecal, which further rejects events with late-developing showers that would pass the trigger selection.  To maintain high efficiency for signal events with recoil electron energies up to 1.2\,GeV, allowing for fluctuations in the electron's reconstructed energy, we require the \ecal energy to fall below 1.5\,GeV.  To model the population of background events that pass this selection, it is crucial to understand the various contributions to the low-energy tail of energy reconstruction.

\begin{figure}[tb]
\centering
\includegraphics[width=10 cm]{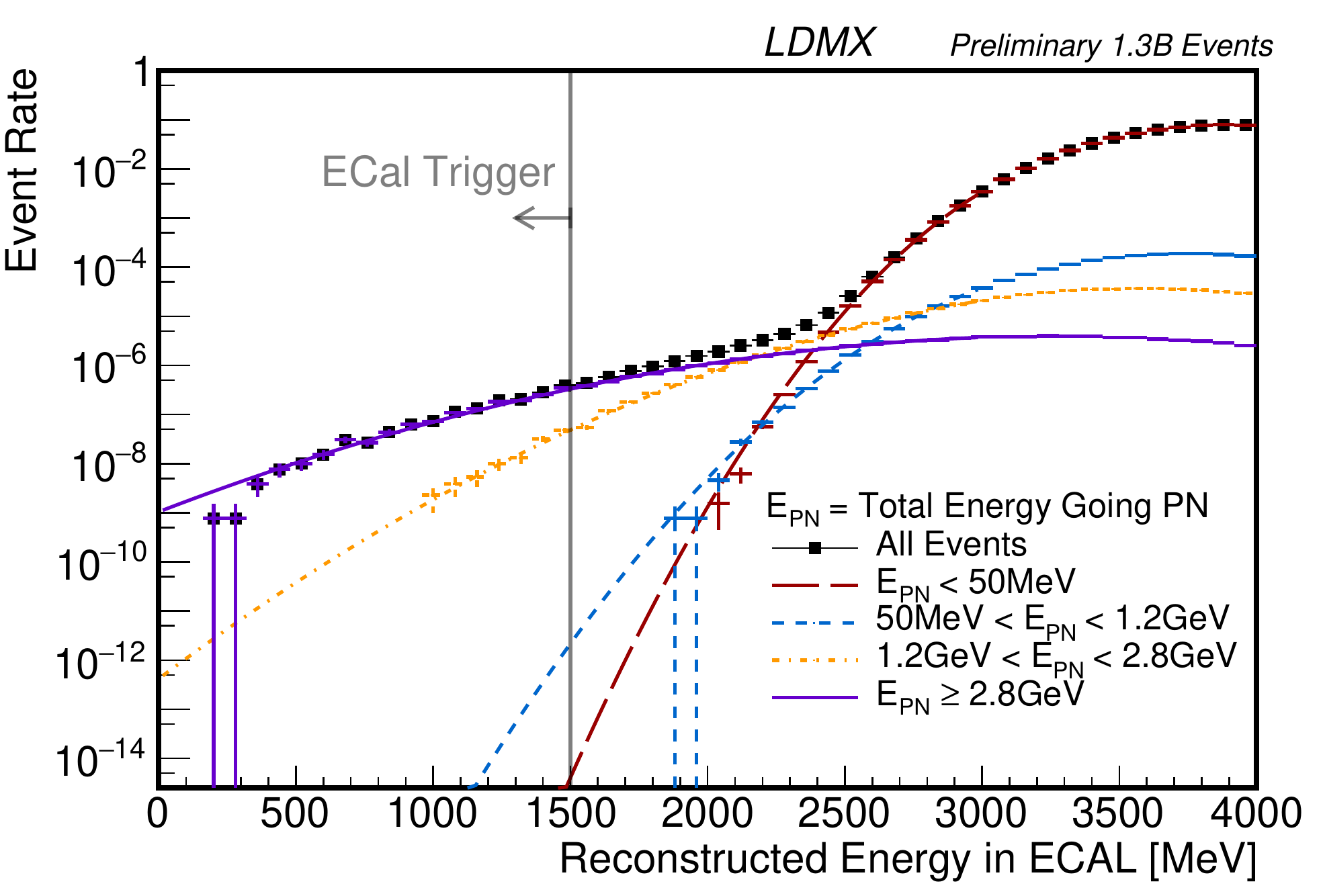}
\caption{
Fraction of events reconstructed with a certain total energy in the \ecal. For each distribution, a selection has been applied according to how much of the energy, $E_{PN}$, is transferred to photo-nuclear reactions (see legend). In all cases, 4 GeV incident electrons are used. The black points show the inclusive sample with no $E_{PN}$ selection applied. The dashed lines are Gaussian fits to the distributions.   The trigger and offline ECal energy requirement select events to the left of the gray line, i.e.~those with reconstructed energy $<1.5$ GeV.
}
\label{fig:em_shower_egamma}
\end{figure}

Figure \ref{fig:em_shower_egamma} shows the spectrum of the energy reconstructed in the \ecal{} for the nearly-inclusive 4 GeV electron event sample described in Sec.~\ref{sec:simSamples}.  The black points show the spectrum for the whole sample, while colored error bars correspond to partitions of the event sample according to $E_{PN}$, defined as the sum of energies of all photons in the event that undergo rare photo-nuclear reactions.  We note that in the vast majority of events, $E_{PN}$ is also approximately the energy of the \emph{hardest} photon undergoing a photo-nuclear reaction. 
The inclusive distribution of energy has a large non-Gaussian tail, but each component sub-divided according to  $E_{PN}$ is well described by a Gaussian fit, as shown by the dashed curves in the figure.  
The maroon points correspond to pure and almost-pure electromagnetic showers (those with less than 50 MeV, or $1.25\%$ of the beam electron's energy, going into photo-nuclear interactions), which dominate the core of the reconstructed energy distribution.   The lower tail of reconstructed energy for such events is very well modeled by a Gaussian fit (dashed maroon curve) over 9 orders of magnitude.  In a $4\times 10^{14}$ EoT run, no such events are expected to survive the 1.5 GeV energy cut.
The distribution of events with 50 MeV to 1.2 GeV (i.e. $1.25\%$ to $30\%$ of the beam electron's energy) going into photo-nuclear reactions (blue error bars) have a discernibly longer tail, but none of these events are expected to survive either, once a further cut on the electron track momentum is applied, as described in the next subsection.  
Instead, the low-energy tail is entirely dominated by events with a hard bremsstrahlung photon with $E>2.8$ GeV (i.e. 70\% of the incident electron's energy, purple error bars).   These reactions can transfer an appreciable fraction of the shower's energy into MIPs and neutral hadrons, giving rise to low reconstructed energy in the ECal. Events with $1.2$ GeV $<E_{PN}< 2.8$ GeV (30\%--70\% of the beam electron's energy transferred to hadrons, yellow error bars) appear as a subleading component.  The vetoes described in the following sections take advantage of the other signatures that these particles produce, in the spatial distribution of energy in the \ecal, particle penetration into the \hcal, and production of tracks in the \ecal by short-ranged MIPs.  

\subsection{Track Selection}\label{ssec:tracksel}
In addition to the energy reconstruction above, we require exactly one track with a momentum of less than 1.2\,GeV.  This selection ensures that the electron lost its energy in the target, rather than in the thicker \ecal.  This requirement suppresses the purple background component in Fig.~\ref{fig:em_shower_egamma} by a factor of 10 and all other components by a factor of 30.  Track multiplicity also provides valuable information.  Events in which the hard bremsstrahlung photon interacts in the target --- whether by ordinary conversion, photo-nuclear scattering, or conversion to muon pairs --- frequently produce multiple tracks.  We also note that the single-track requirement plays an important role in rejecting rare reactions initiated by the electron (the left branch of Fig.~\ref{fig:BackgroundsChart}, not discussed in detail in this paper): for example, charged-current neutrino production typically has no outgoing track, while electronuclear reactions typically produce multiple tracks.  

After the cut on missing energy and the requirement of a single track with $p<1.2$ GeV in the recoil tracker, the dominant source of background is events where an electron undergoes a hard bremsstrahlung reaction in the target, and the produced photon undergoes a photo-nuclear  or muon-conversion reaction in the target or \ecal.  
These are the event classes for which biased samples, at equivalent statistics of $2 \times 10^{14} - 2 \times 10^{15}$ EoT, have been produced (see Table \ref{tab:bkg_samples}).  
In the following subsections, we introduce several additional selections that efficiently reject these backgrounds.  In discussing their performance, we focus on the largest and most difficult to reject of the four samples, namely ``\ecal PN'' events, in which the bremsstrahlung photon undergoes a photo-nuclear reaction in the \ecal.  

\subsection{Boosted Decision Tree Based on \ecal Features}
\label{ssec:bdt}

Several variables constructed from \ecal information have been found to be powerful in rejecting background reactions. 
These were studied in the context of PN reactions occurring in the \ecal,
but they are also effective in rejecting target PN events as well as muon conversions, as we will see below. 

\subsubsection{Global Features}
\begin{figure}[tb]
\centering
\includegraphics[width=0.4\textwidth]{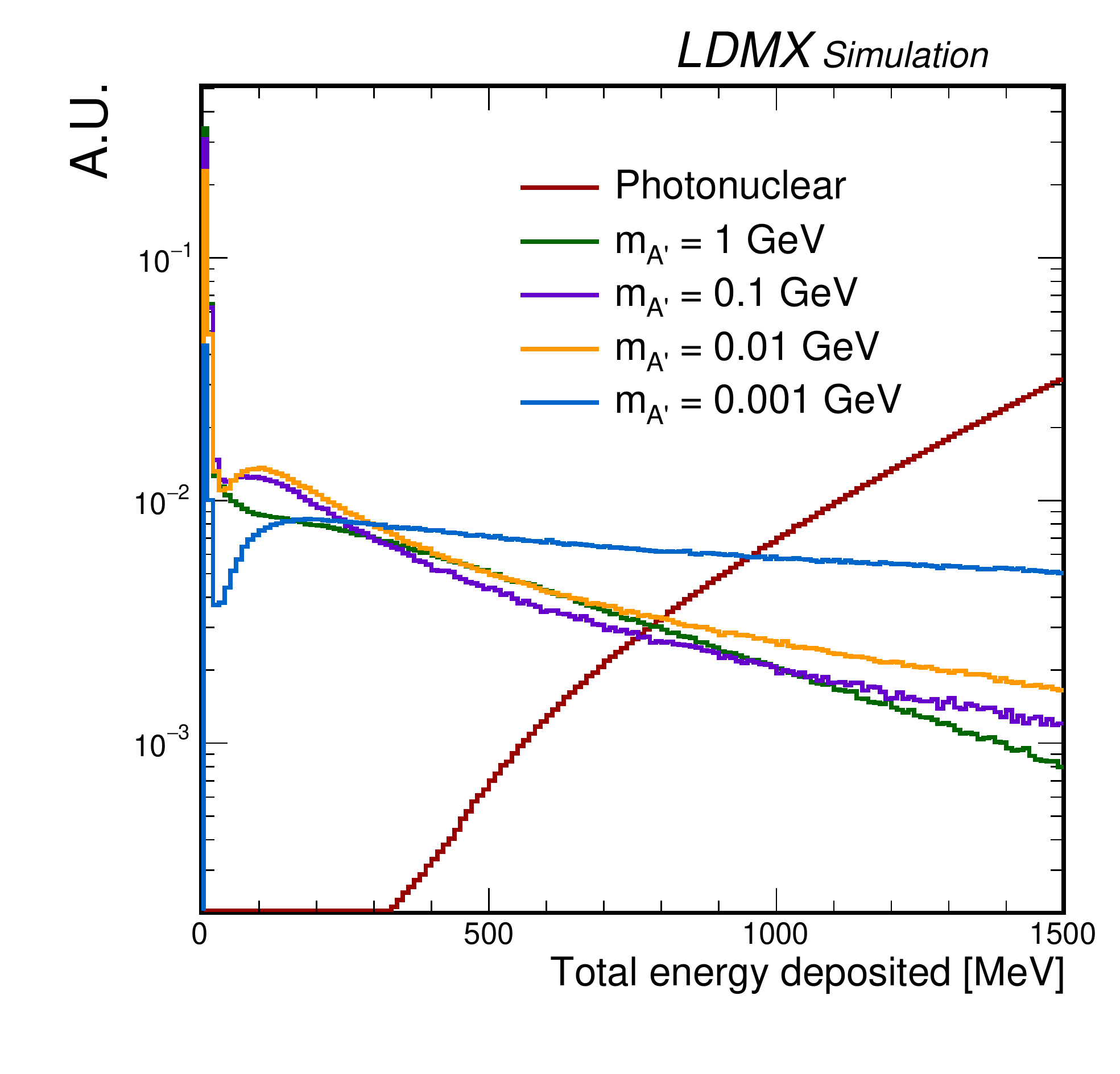}
\includegraphics[width=0.4\textwidth]{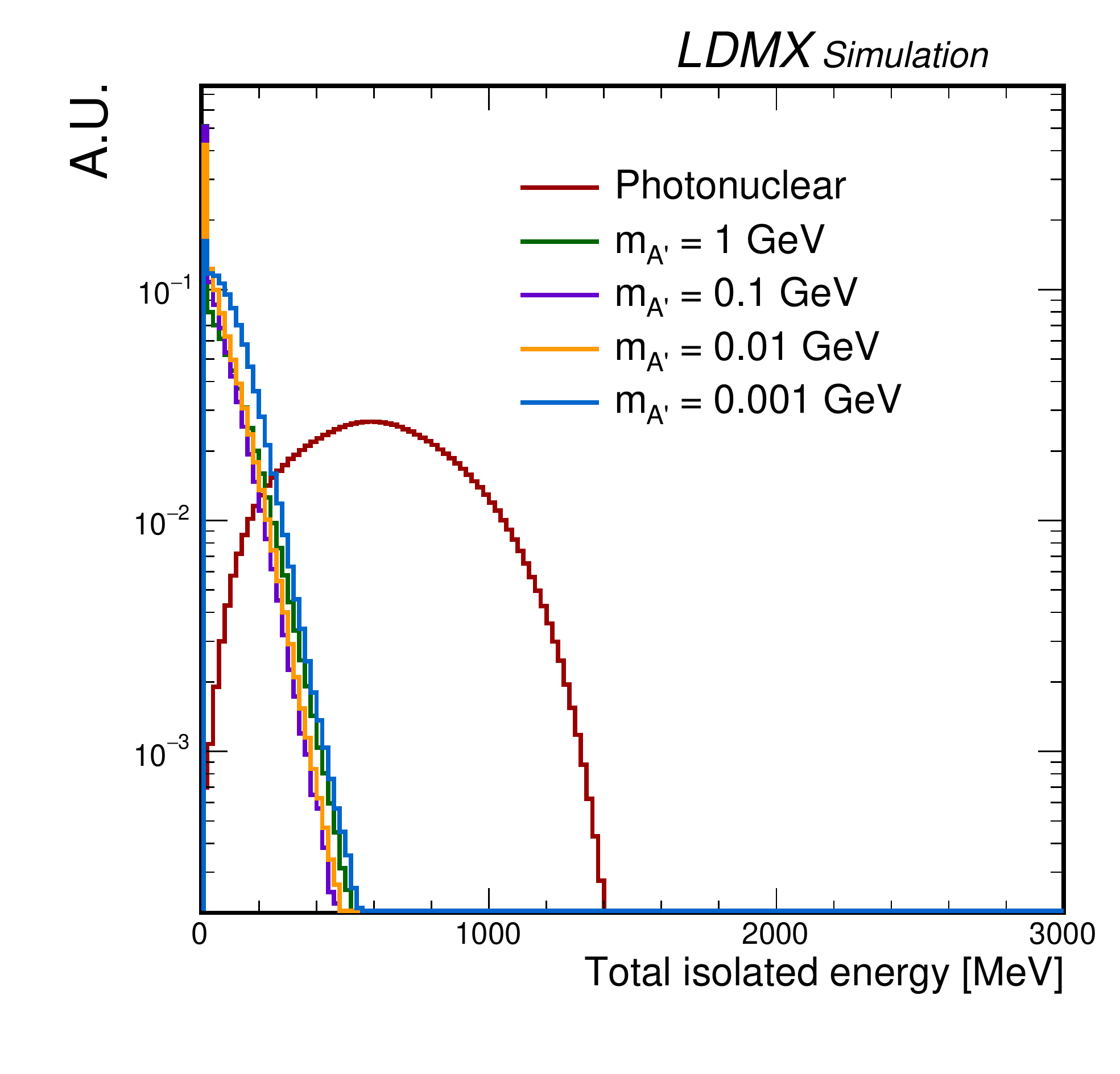}\\
\includegraphics[width=0.4\textwidth]{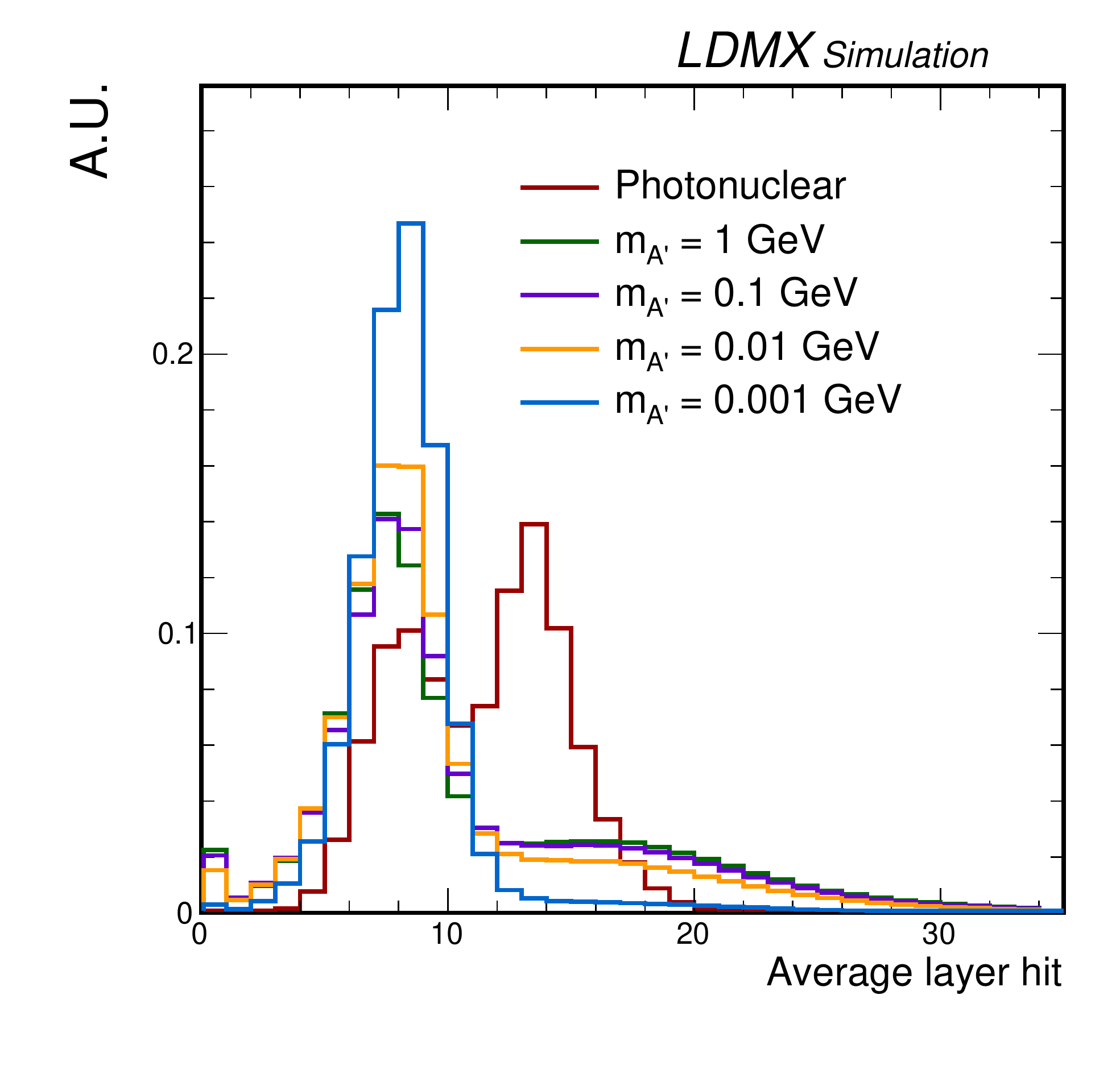}
\includegraphics[width=0.4\textwidth]{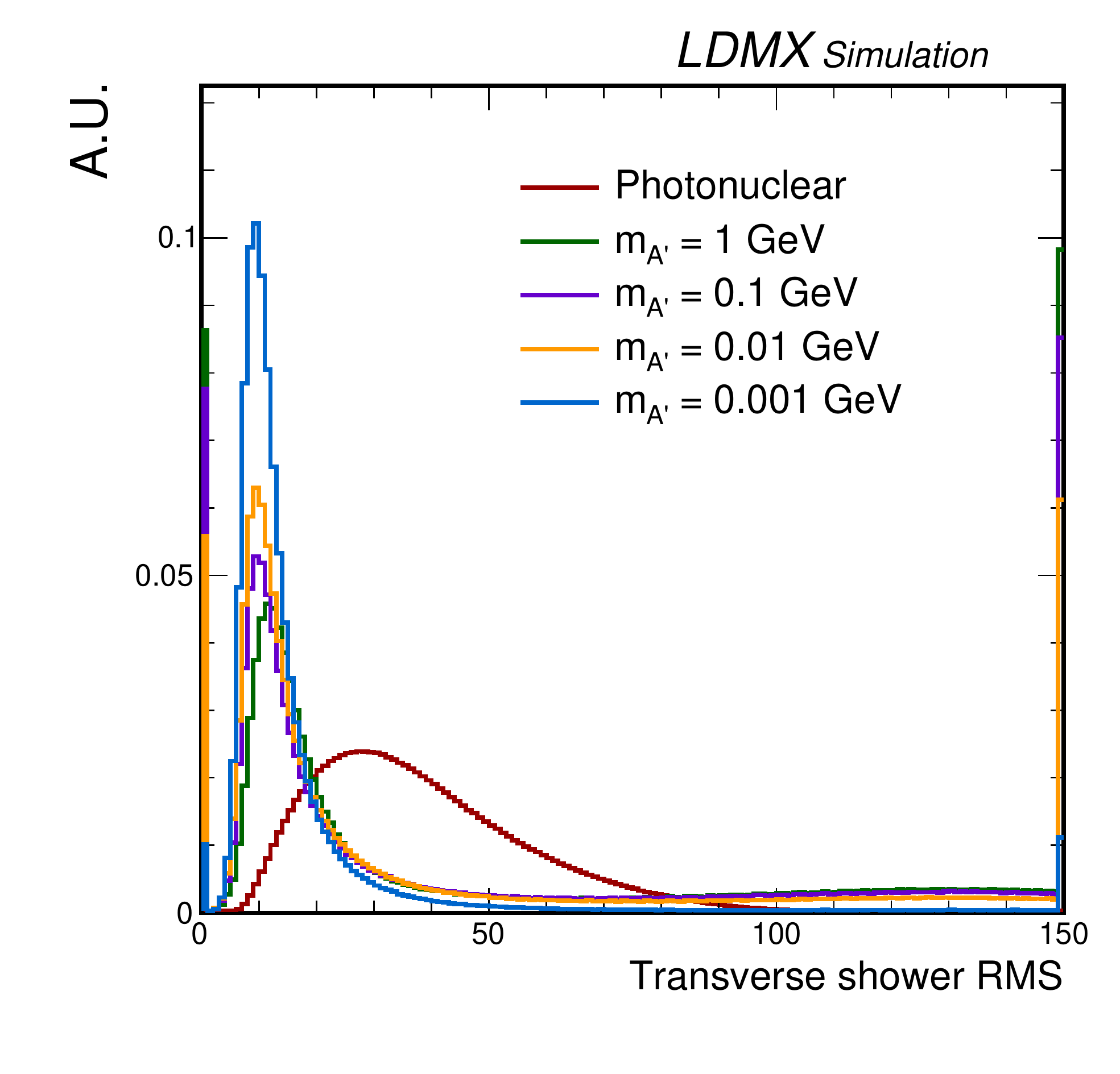}
\begin{flushleft}
\caption{\label{fig:ecalvars} Distributions of quantities related to the energy deposited in the \ecal for photo-nuclear and signal processes in events passing the trigger. 
From top left to bottom right: total reconstructed energy, total isolated energy, energy-weighted average layer number \LL, energy-weighted transverse RMS. The distributions are shown for events in which the total energy reconstructed in the \ecal is less than 1.5\,GeV. All distributions are normalized to unit area. For more detail about the definition of these variables, we refer the reader to the corresponding text. 
}
\end{flushleft}
\end{figure}

As a first step, we consider \emph{global} \ecal variables, i.e. features like sums or averages over the entire \ecal{}. 
Such variables already provide efficient signal and background separation, as illustrated with the examples in Fig.~\ref{fig:ecalvars}. 
%
%
In the vast majority of PN interaction background events, both the recoil electron and the products of the PN reaction deposit energy in the \ecal, such that the overall energy deposition, shown on the top left,  
is generally larger than what is found for signal. 
The top right in Fig.~\ref{fig:ecalvars} shows the total isolated energy, computed as follows. 
Cells with hits above the readout threshold are considered to be isolated if they have no immediate neighbors in the same layer containing a readout hit. The cell containing the global shower centroid position in a layer and its immediate neighbor cells are not considered. The total isolated energy is the sum of the energy deposited in all isolated cells from all layers.
There is typically little isolated energy for the signal, where there is only one electromagnetic shower from the recoil electron, whereas for the PN background, there are additional hits outside of this shower, distributed more widely, leading to higher values for the isolated energy. 
The peaks near zero in the signal distributions of total energy and total isolated energy (top panels of Fig.~\ref{fig:ecalvars}) are populated by events in which the recoil electron is deflected at large angles and does not enter the \ecal, resulting in very little energy in the \ecal.

Further discriminating power between signal and PN backgrounds can be obtained by exploiting information about the shower profile in the longitudinal and transverse directions in the \ecal. 
As mentioned earlier, PN events are typically characterised by a higher level of activity in the back part of the \ecal compared to signal events. 
One variable exploiting this difference in the \emph{longitudinal} energy distribution is 
the energy-weighted average layer number \LL\, computed from the energy sums of all readout hits in all 34 layers of the ECal, with the first layer being layer 0. The distribution of \LL\, is shown in the bottom left of  Fig.~\ref{fig:ecalvars}. 
The tails in the signal \LL\, distributions come from events where the recoil electron  misses or grazes the \ecal{}, so that the average is dominated by noise hits.  

The \emph{transverse} energy profiles also differ between signal and PN events.  Two effects broaden the transverse profiles of PN events.  First, the energy depositions resulting from the PN reaction have a broader transverse profile than the shower from the recoil electron.  Second, the magnetic field separates the recoil electron from the photon, which also enlarges the region over which energy is deposited. 
The bottom right plot in Fig.~\ref{fig:ecalvars} shows as an example the transverse RMS, defined as the energy-weighted RMS of the transverse distance of all hits from the position of the shower centroid. 
The peak at zero in these distributions is again due to the small fraction of signal events where the recoil electron grazes or misses the \ecal{}.
Distributions of further global variables used as inputs to the BDT are collected in Figs.~\ref{fig:ecalvars_app} and ~\ref{fig:bdtTransLong_app}  in Appendix~\ref{sec:appEcalVar}.


\subsubsection{Shower Containment Variables}
The high density and granularity of the \ecal{} make it possible to obtain good separation between two electromagnetic showers. 
In order to exploit this capability for discrimination between signal and (PN) background, we define regions expected to contain significant fractions of the shower energy originating from the recoil electron, as well as analogous containment regions around the inferred path of the bremsstrahlung photon (assuming the background hypothesis). 

\begin{figure}[tb]
\centering
\includegraphics[width=0.53\textwidth]{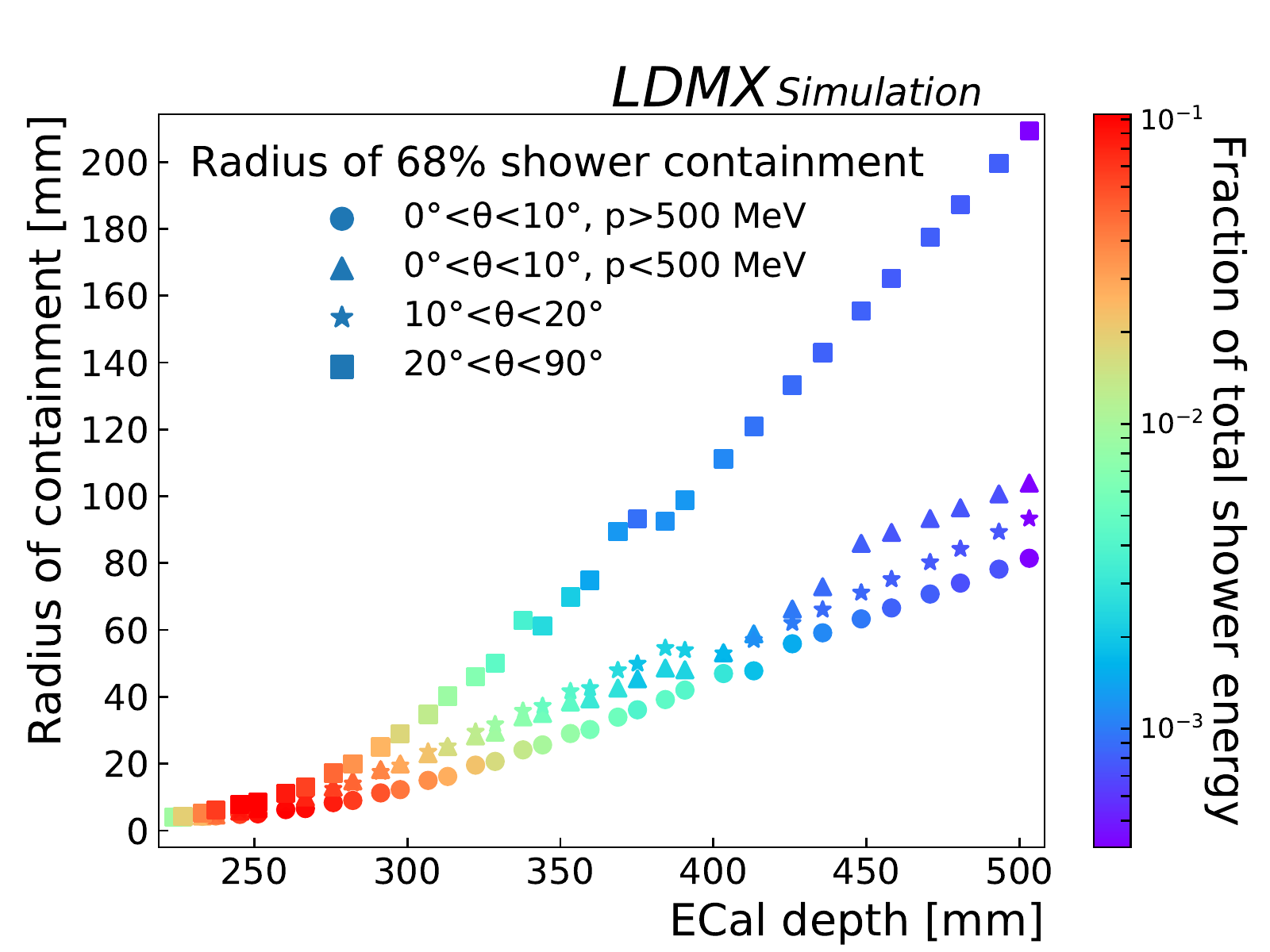}
\includegraphics[width=0.4\textwidth]{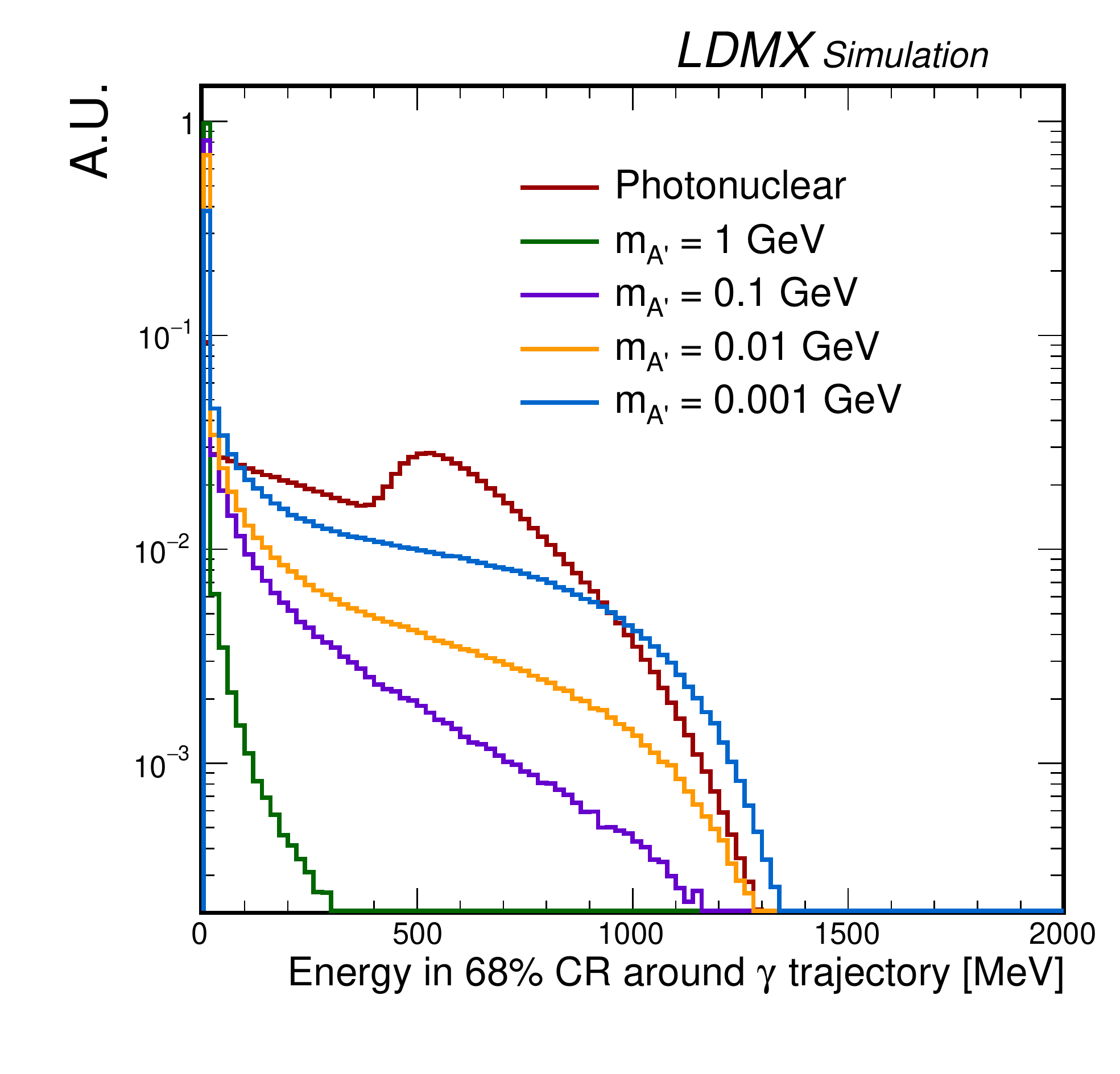}
\begin{flushleft}
\caption{\label{fig:radiusOfContainment} Left: Radii containing 68\% of the energy deposited in individual \ecal{} layers by an electron shower, as a function of depth. The radii are determined for different regions in the phase space of momentum vs angle of incidence at the face of the \ecal. Right: Energy reconstructed in the 68\% containment radius around the projected photon trajectory for PN background and signal events (distributions are normalized to unit area).}
\end{flushleft}
\end{figure}

The radial containment of electromagnetic showers is determined from simulation as plotted in Fig.~\ref{fig:radiusOfContainment}, left, which shows the radius of the circular region in each \ecal{} layer that on average contains 68\% of the total electromagnetic shower energy deposited in that layer. Since the profile and shape of the shower generated by a recoil electron depend on its momentum and the angle at which it enters the \ecal{}, the 68\% containment radius is determined separately for different ranges of these variables. Based on these radii, \emph{containment regions} are defined around the projected trajectories of the electron and, treating the event as background, also around the inferred photon direction as described in more detail in Appendix~\ref{sec:appEcalVar}.

For signal, little to no energy is expected to be seen in the photon containment regions or the region outside the electron and photon containment regions. The lower the dark photon mass, however, the less the electron is deflected, resulting in overlap between the electron and photon regions and hence some energy deposited by the recoil electron in the photon containment region. The corresponding distribution for th PN background is broader and extends to higher values, allowing fo discrimination between signal and background. As an example, Fig.~\ref{fig:radiusOfContainment}, right, shows the distribution of the energy within the containment radius around the photon trajectory. Further related distributions are collected in Fig.~\ref{fig:containmentVars_app} in Appendix~\ref{sec:appEcalVar}.


\subsubsection{Performance of the Boosted Decision Tree} 
\label{sss:bdt}
The containment variables, as well as the discriminating variables sensitive to the global transverse and longitudinal profiles and the energy-related variables described previously, are exploited in a multivariate analysis based on a BDT. 
This BDT is trained on PN background events and a mixture of signal events at the four mediator mass hypotheses listed in Sec.~\ref{sec:simSamples}.

\begin{figure}[tb]
\centering
\includegraphics[width=0.48\textwidth]{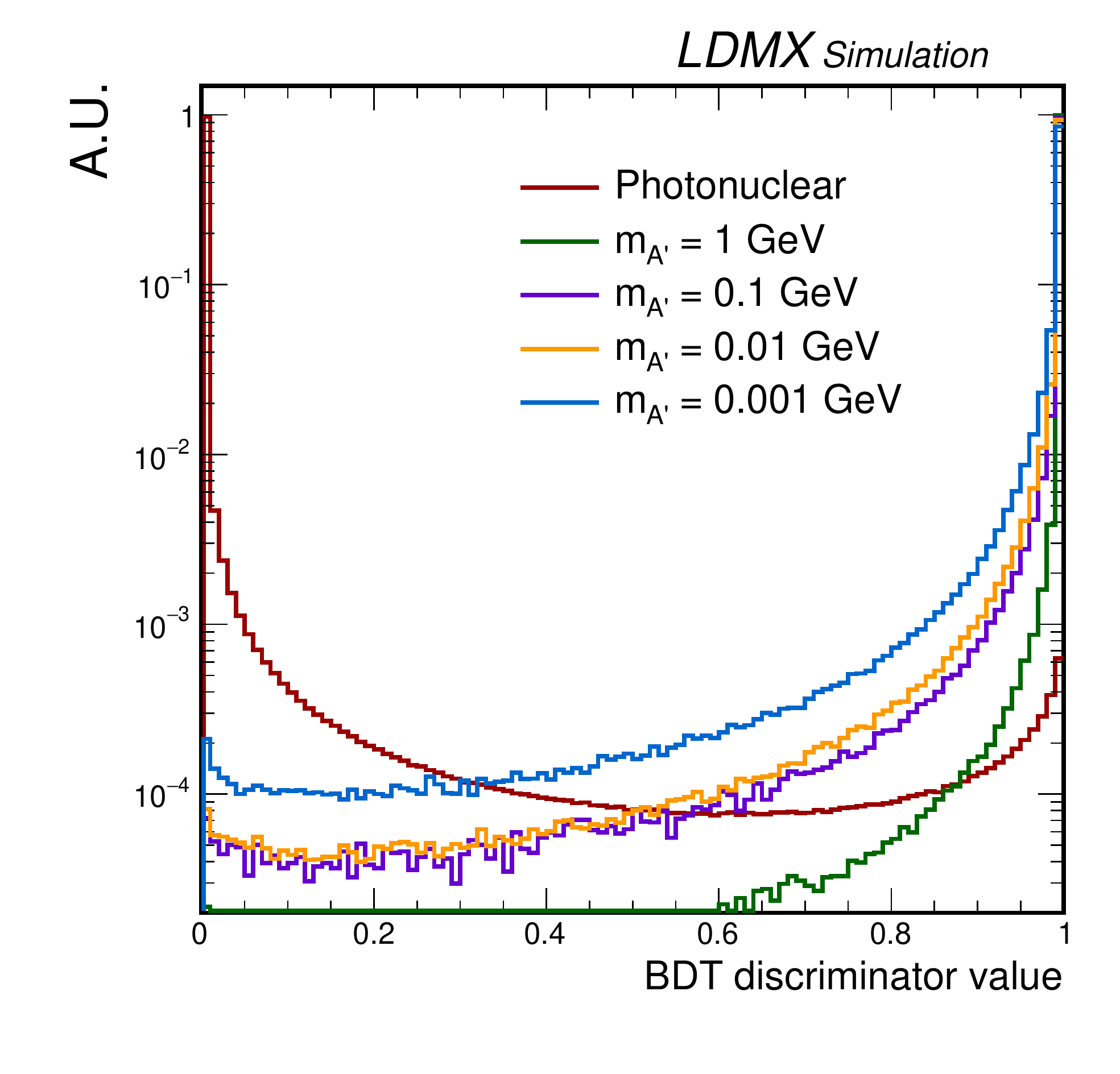}
\includegraphics[width=0.48\textwidth]{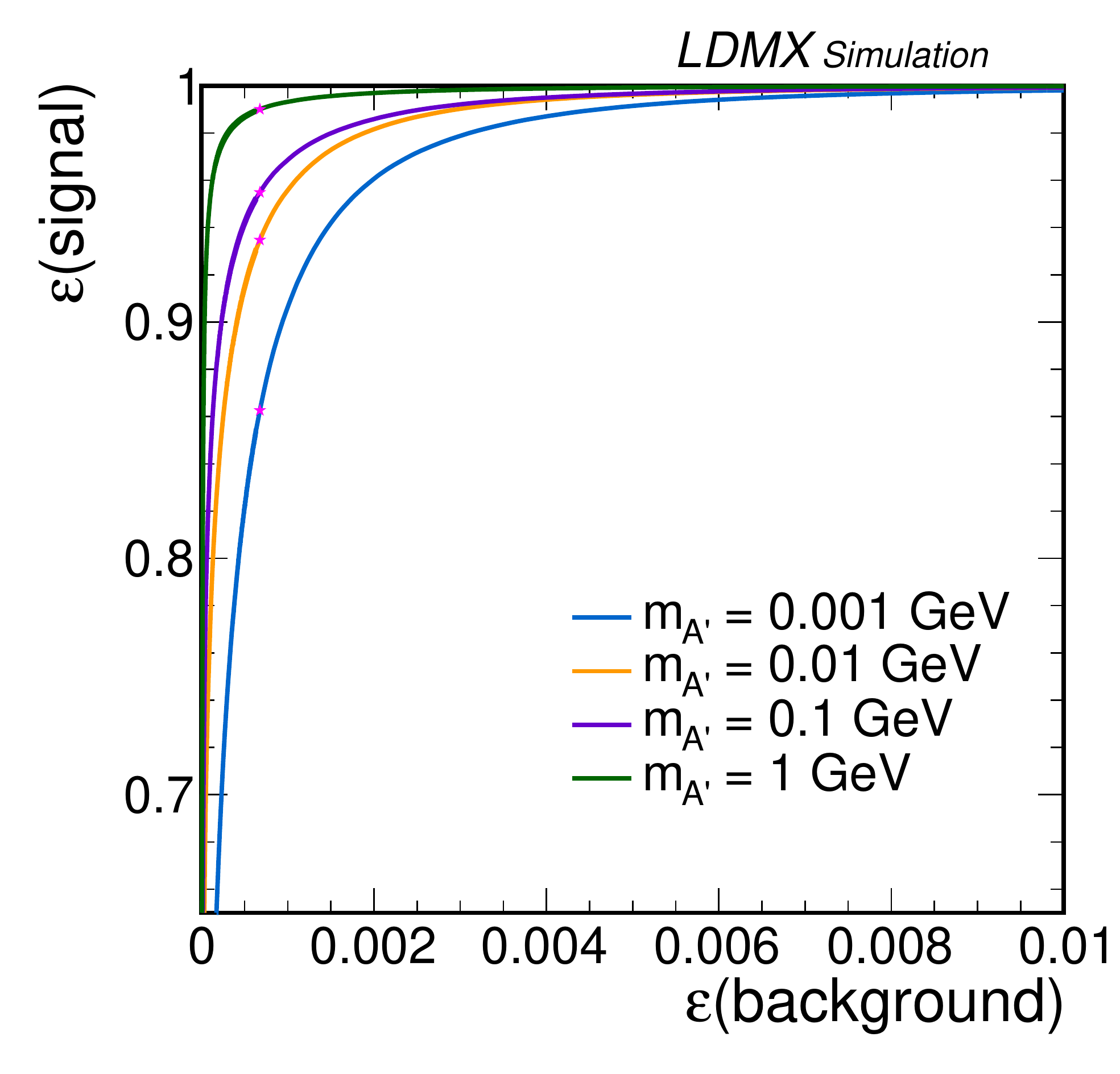}
\begin{flushleft}
\caption{\label{fig:bdt} Left: distributions of the \ecal BDT discriminator value for signal and photo-nuclear events passing the trigger. All distributions are normalized to unit area. Right: signal efficiencies as a function of the background efficiency obtained by varying BDT thresholds for the \ecal BDT evaluated for signal and photo-nuclear background events passing the trigger. The pink markers indicate the efficiencies corresponding to a discriminator threshold of 0.99 in this sample.}
\end{flushleft}
\end{figure}

Figure~\ref{fig:bdt} (left) shows the distributions of the BDT discriminator values for signal and background events selected by the trigger. 
The signal vs. background efficiency obtained by varying discriminator thresholds is drawn in Fig.~\ref{fig:bdt} (right) for each mediator mass considered. 
The pink markers indicate the efficiencies in each sample corresponding to a discriminator threshold of 0.99.
Requiring the BDT value to be greater than 0.99 retains approximately 85-99\% of the signal, depending on the mediator mass, while rejecting more than 99.9\% of photo-nuclear reactions occurring in the \ecal.  

The same BDT and discriminator cut is used for all of the background samples.
For muon conversions in the \ecal and target, the BDT requirement rejects all of the remaining events.   
Here, the BDT veto is particularly useful for events that contain one soft muon that does not leave the \ecal, and one hard muon that decays into an electron and neutrinos, since such events are hard to veto by the other detector systems.   
The BDT also rejects 95\% of target PN events surviving the preceding vetoes.

\subsection{HCal Veto}\label{ssec:hcalveto}
The energy deposited in the \hcal offers a complementary veto for photo-nuclear reactions.  For the present study, we define a vetoable hit to be a single scintillator bar in which at least five photoelectrons are produced in-time with the beam electron. 
The \hcal veto requires there to be no such hit in the entire \hcal, otherwise the event is rejected.  The veto threshold of five PEs is chosen to  minimize the signal inefficiency due to the noise.  The rejection capability is not, however, especially sensitive to this threshold.  

\begin{figure}[tb]
\centering
\includegraphics[width=0.99\textwidth]{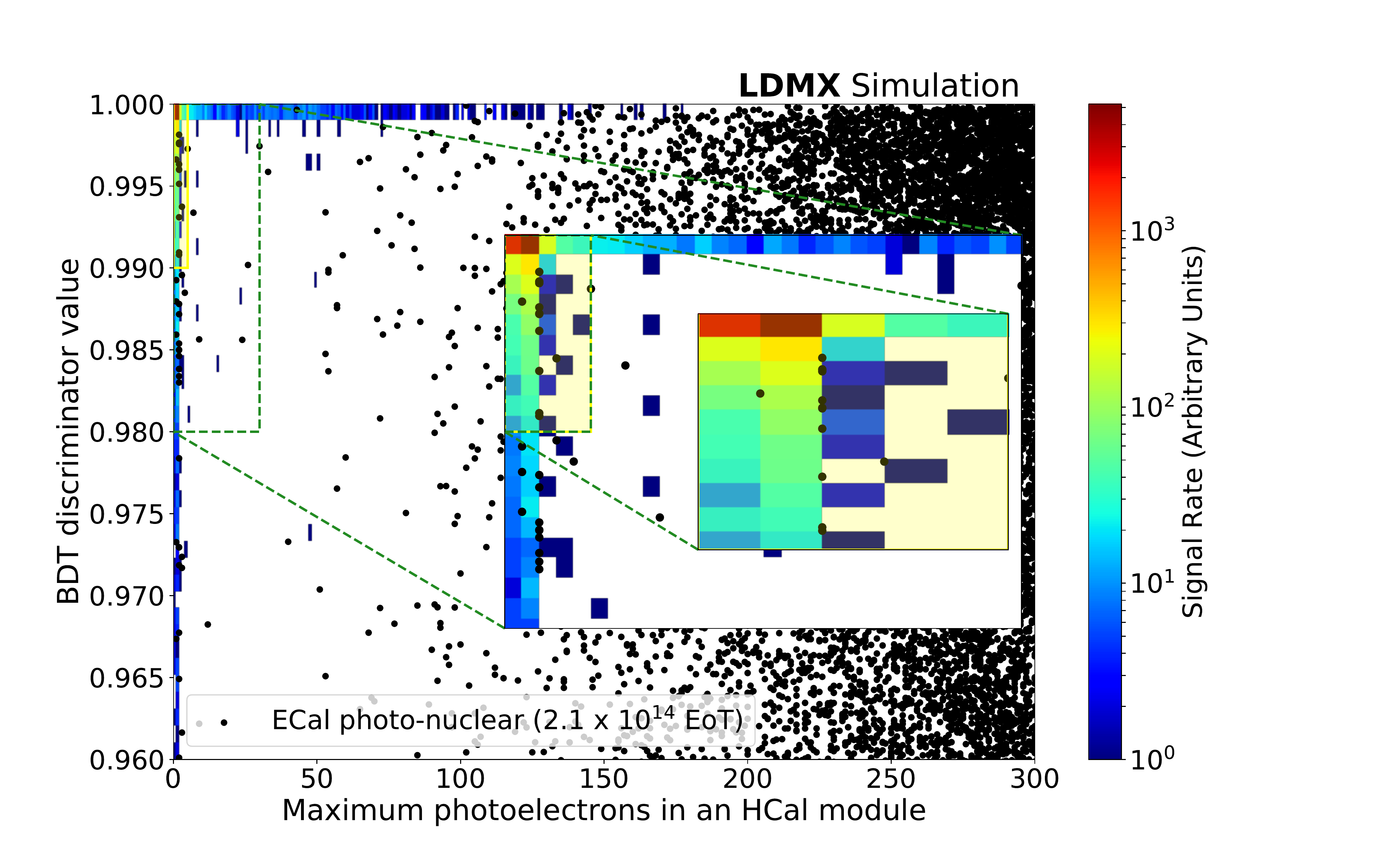}
\begin{flushleft}
\caption{\label{fig:bdt_hcal} Distribution of the \ecal BDT discriminator value ($y$ axis) and 
maximum number of photoelectrons (PEs) in any \hcal module ($x$ axis) for an ECal photo-nuclear 
background sample (black) equivalent to $2.1 \times 10^{14}$ electrons on target.  A representative 100 MeV, 
$A'$ signal sample is also shown as a heatmap. The signal
band at large max PE is populated by 
events where the recoil electron is produced softly,  
misses the \ecal and showers in the side \hcal.  The signal band at 
low max PE is composed of events where the recoil 
electron shower is fully contained in the \ecal. In the analysis, the signal region (yellow box) is 
defined by events with a BDT score $<0.99$ and an a maximum number of PEs in an HCal module 
of $<5$. As is evident from the figure, a majority of the signal lies within the defined signal
region.  The background events within the signal region are rejected by additional requirements
on the tracks in the Recoil tracker and the ECal.}
\end{flushleft}
\end{figure}

Figure \ref{fig:bdt_hcal} illustrates the combined power of the \ecal BDT and HCal veto for background rejection.  The horizontal axis of the figure is the maximum number of photoelectrons in any HCal bar, in time with the event.   The signal region defined by the previously described ECal BDT and HCal PE cuts
is also shown in yellow.  Typical photo-nuclear events are vetoed by both the \ecal and the \hcal, and many of these are vetoed by the \hcal by a wide margin (e.g. a typical hadronic shower will leave a large number of hits, some of these with hundreds of PEs in individual hits; a muon penetrating into the \hcal will also leave a large number of hits well above the threshold).  
However, there are several types of events where the \ecal has limited rejection capability.  The majority of events surviving the \ecal BDT veto arise from hard bremsstrahlung photons that undergo photo-nuclear reactions in the target area. Events where softer charged secondaries are deflected out of the \ecal geometric acceptance and into the \hcal survive the \ecal veto.  They are, however, readily rejected by the \hcal since they typically have multiple final state particles that deposit energy in the \hcal through ionization even if they do not shower.  

The second class of events not rejected by the \ecal BDT are those where most of the photon's energy is carried by a single neutral hadron (neutron or $K^0_L$).  Because neutral hadrons can escape the \ecal without depositing a significant amount of energy, these events dominate the sample of \ecal PN events surviving the BDT veto.  
The \hcal geometry has been optimised to detect these neutral hadrons with high efficiency. 

Applying the \hcal{} veto rejects 99.9924\% of the \ecal PN background remaining after the BDT cut, while preserving 99\% of the signal.  
This selection applied to PN reactions from the target rejects all of the events passing the BDT selection.

\subsection{Track Features in the \ecal}\label{ssec:trackveto}
After all these cuts, the remaining background arises mostly from PN reactions resulting in a single charged kaon. 
As discussed earlier, the kaon may decay in flight within the \ecal, transferring most of its energy to a neutrino and leaving only a short track within the \ecal. 
Such event signatures are not detectable in the \hcal and are difficult to distinguish from signal for the BDT. 
However, the high granularity of the \ecal provides possibilities to track MIPs as they pass through the \ecal. 
Thus, as a final step of this analysis, a tracking algorithm has been introduced to identify tracks left by charged particles originating from photon interactions in the \ecal.  These selections were developed and tuned on a sample of $\sim10\times10^{3}$ \ecal PN events surviving the preceding \ecal and tracker vetoes, but \emph{prior to applying the \hcal veto}, then applied to the ten events that were not vetoed by the \hcal.

In a first stage, tracks normal to the back of the \ecal are formed from combinations of hits in cells directly in front of each other and not more than two layers apart. 
The second stage uses a linear regression among certain three-hit combinations of the remaining hits. 
At both stages, tracks are discarded if they are too far from the projected photon trajectory or too close to the projected electron trajectory. 

\begin{figure}[htp]
\centering
\includegraphics[width=0.49\textwidth]{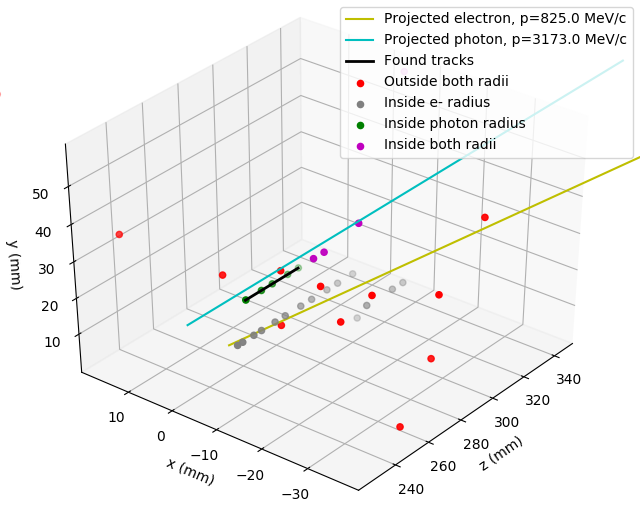}
\includegraphics[width=0.46\textwidth]{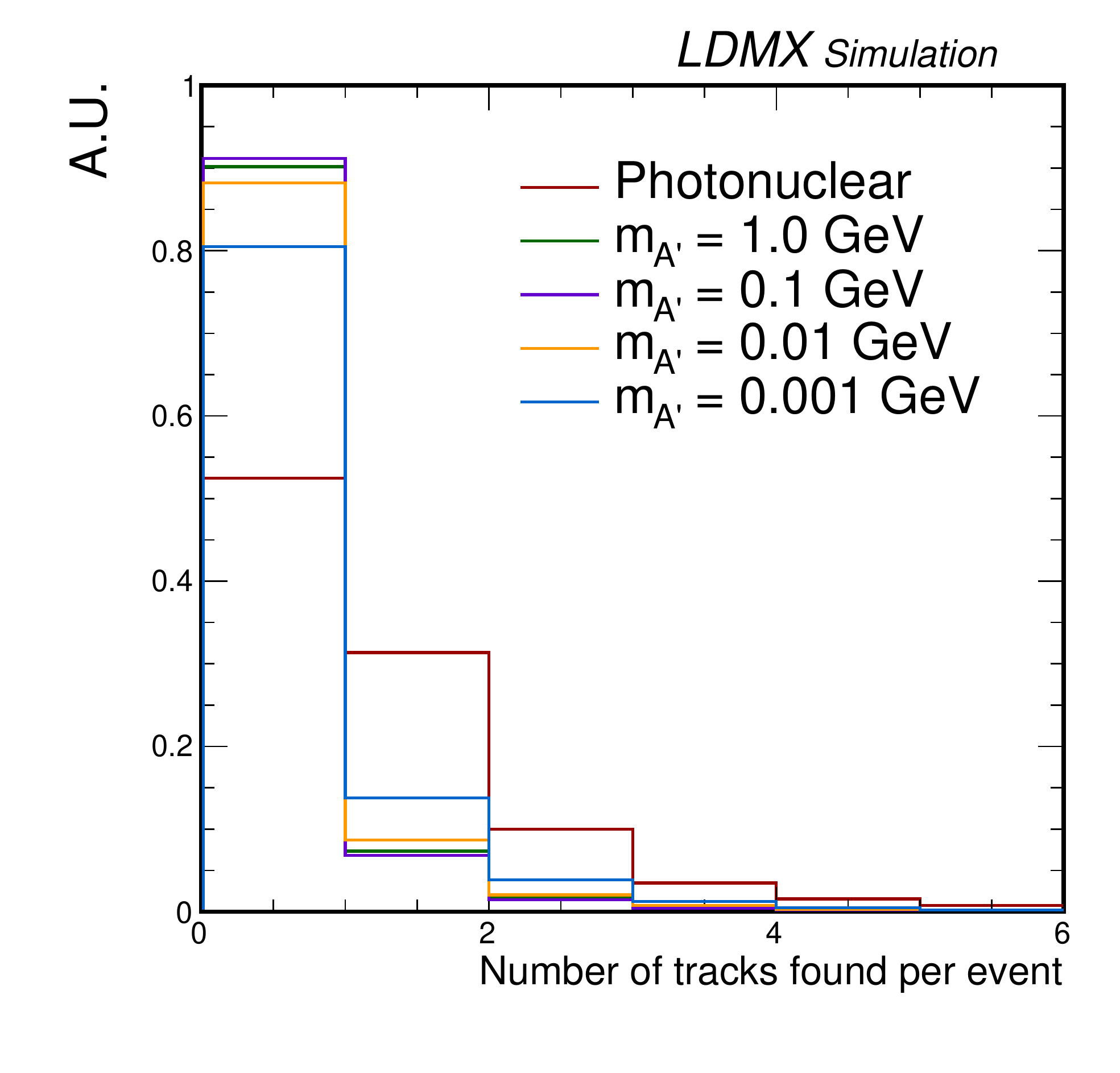}
\caption{\label{fig:MIPtrk}Left: visualization of a simulated PN event that survives both the \ecal{} BDT and \hcal{} activity vetoes, but is rejected by the \ecal{} tracking algorithm. Hits are color-coded according to whether they are inside or outside the electron and photon containment radii, and the black line indicates the reconstructed track. Right: number of \ecal{} tracks detected per event in a sample of 10,000 events each for signal and PN background after \ecal vetoes; normalized to unit area.}
\end{figure}

Figure~\ref{fig:MIPtrk} (left) shows a visualisation of one of the background events surviving the previous selections to which the track finding has been applied, resulting in the track shown in black, close to the projected photon trajectory in cyan. 
The right plot in Fig.~\ref{fig:MIPtrk} shows the distribution of the number of tracks found by the methods described above in signal and PN background events. Any event with one or more tracks is rejected. 

A final criterion is useful for identifying background events in which no track is produced, but in which potentially isolated hits are present in the vicinity of the photon trajectory in the early \ecal{} layers. If any hits that are outside the electron radius of containment are found to be within one cell of the photon trajectory in the first 6 layers of the \ecal, the event is rejected. 
The full chain of these cuts rejects all remaining background events, at a moderate cost in signal efficiency.

\FloatBarrier

%% file: sections/ResultsDiscussioin.tex
The number of background events left after each selection stage as described in Sec.~\ref{sec:evtSel} is summarized in Table~\ref{tab:bkg_rejection}. In this table, the trigger selection and the cut on the total energy reconstructed in the \ecal have been combined. 
The signal efficiency for the same sequence of cuts varies between approximately 30\% and 50\% in the mediator/pair mass range from 1\,MeV to 1\,GeV. 

\begin{table}[!t]
    \centering
    \begin{tabular}{ l | c | c | c | c  } 
        
        \hline\hline
            & \multicolumn{2}{c|}{\textbf{Photo-nuclear}}   & \multicolumn{2}{c}{\textbf{Muon conversion}} \\ \cline{2-5}
                                       & \textbf{Target-area}  & \textbf{\ecal}        &  \textbf{Target-area} & \textbf{\ecal}        \\ [0.5ex]
        \hline
        EoT equivalent                          & $4 \times 10^{14}$ & $2.1 \times 10^{14}$  & $8.2\times10^{14}$   & $2.4\times10^{15}$    \\ 
        \hline
        \hline
        Total events simulated                  & $8.8 \times 10^{11}$  & $4.65 \times 10^{11}$ &  $6.27\times10^{8}$   & $8\times10^{10}$    \\
        Trigger,  ECal total energy $<1.5$ GeV  & $1 \times 10^{8}$     & $2.63 \times 10^{8}$  &  $1.6\times10^{7}$    & $1.6\times10^{8}$     \\
        Single track with $p<1.2$\,GeV & $2 \times 10^{7}$     & $2.34 \times 10^{8}$  &           $3.1\times10^{4}$           & $1.5\times10^{8}$     \\
        \ecal BDT ($>0.99$)            & $9.4 \times 10^{5}$   & $1.32 \times 10^{5}$  &  $<1$                   & $<1$                     \\
        HCal max PE $<$ 5              & $<1$                     & 10                    &  $<1$                   & $<1$                     \\
        \ecal MIP tracks = 0               & $<1$                     & $<1$                     &  $<1$                   & $<1$                     \\
        \hline
    \end{tabular}
    \caption{The estimated levels of photo-nuclear and muon conversion backgrounds after
             applying the successive background rejection cuts outlined in this paper.  
             Here, the total events simulated corresponds to the total electrons fired
             on target in the simulation. The biasing factor passed to the \geant 
             occurrence biasing toolkit is used to scale the total events simulated 
             to the electron on target (EoT) equivalent.}
    \label{tab:bkg_rejection}
\end{table}

%

Muon conversion backgrounds in LDMX are expected to be rejected completely by the selections on tracks and energy depositions in the \ecal{}, even for a number of EoT far beyond the initial $4\times10^{14}$ EoT run at 4 GeV that is currently envisioned. No target-area PN events remain after the BDT selection for $4\times10^{14}$ EoT. 
The statistics of the \ecal PN simulation used in this analysis corresponds only to $2\times10^{14}$ EoT, but given that all of these events are rejected by the full analysis chain,  
at most a few events could be expected for the full initial run. 
As was illustrated in Fig.~\ref{fig:reach1}, this would reduce the reach in the low-mass region, but still allow to probe beyond several thermal targets.  

However, further development of the analysis techniques should make it possible to improve the background rejection in the future.  Some of the improvements being explored are refinements of the cutflow presented above, using the same basic information from the detectors --- for example, extending the HCal veto to reject clusters of nearby hits that are individually below the 5PE threshold, integrating the track-identification into the feature-based BDT, and adding longitudinal segmentation to the shower containment variables in the BDT.  

A second powerful tool for rejecting photo-nuclear backgrounds is the reconstruction of the transverse momentum of the recoil electron.  The striking difference in transverse momentum distributions between DM signals and photo-nuclear background was shown in Fig.~\ref{fig:LDMspectra} (right), but is not exploited at all in the  analysis presented above.  These are generator-level distributions, but smeared for multiple scattering in the target, which is the leading source of $p_T$ uncertainty for photon-induced backgrounds.  Not using $p_T$ reflects a deliberate choice: LDMX's design goal is to reject background to the $O(1)$-event level \emph{before} any transverse momentum cuts, enabling optimal sensitivity to DM pairs as light as $\sim 1$ MeV (where transverse momentum is a poor discriminator) and allowing the use of $p_T$ as a final cross-check and as a means of characterizing signal events.  If backgrounds are unexpectedly high, however, exploiting the discriminating power of transverse momentum will allow LDMX to maintain excellent sensitivity over most of the DM mass range of interest, as shown in Fig.~\ref{fig:reach1}.    

This study addresses the rejection capabilities of LDMX in Monte Carlo. The LDMX physics program will also require a data-driven validation of both the physical rates and kinematics of key background reactions and the detector performance.  The veto possibilities introduced in Sec.~\ref{sec:evtSel} will allow reconstruction of rates and kinematics for limiting background topologies and closely related reactions. For example, kaon decays in flight with a short but visible track are far more numerous than those that leave no reconstructed track, and can be used to measure the rate of single kaon production.  The variety of complementary veto opportunities is also valuable in characterizing the detector's inefficiency in rejecting these reactions.  There are numerous control samples available in data, accessed, for example, by inverting or loosening one of the selections, studying events with lower-energy PN photons, or considering only events vetoed within a smaller geometric region of the \hcal.  
Finally, events with low recoil electron $p_T$ define a control region where hard bremsstrahlung is enriched relative to signal, completely independent of the instrumental rejection capability.  The veto efficiencies of Sec.~\ref{sec:evtSel} are independent of $p_T$ to within $\sim 10\%$, so that veto efficiencies at low $p_T$ can be reliably extrapolated to a higher-$p_T$ signal region.

The beam energy of 4\,GeV considered in this analysis is a realistic scenario for an initial run, but the full statistics of LDMX will likely be collected at a significantly higher beam energy. Higher beam energies that will become available at SLAC (8\,GeV from LCLS-II HE \cite{Raubenheimer:2018wwc}) or potentially at CERN (up to 16\,GeV, \cite{Akesson:2019drp}) will further improve the sensitivity for several reasons. On the one hand, the signal cross sections increase, improving the sensitivity, particularly in  the high mass region (several hundred MeV). On the other hand, the rates of certain backgrounds decrease with higher energy, e.g. that of the exclusive 2-body PN reactions scales as $E_{\gamma}^{-3}$, and the products from these reactions carry more energy and are hence more visible in the detector. Similarly, in-flight decays within the detector, e.g. of charged kaons, have a lower rate and more detectable products at higher energy. One important potential background that does increase in rate with increasing beam energy is neutrino production. But because this background has a yield below the $10^{-3}$ level for a $4\times10^{14}$ EoT run at 4 GeV, roughly O($10^{17}$) EoT can be tolerated at even 8 GeV beam energy before this background becomes problematic. 




%% file: sections/Conclusion.tex
The existence of thermal-relic dark matter in the sub-GeV mass region is well motivated and its exploration an important goal of the community.  
Accelerator-based experiments like LDMX are an irreplaceable key-component of the global search program. For the fixed-target missing-momentum approach unique to LDMX, background events are dominated by photon-induced reactions, and their rejection is a benchmark for the performance of the experiment. Particularly challenging event types are photo-nuclear reactions resulting in single neutral or charged particles, that can be rejected thanks to the efficient hadronic calorimeter and/or the highly-granular electromagnetic calorimeter. Based on detailed \geant-based full simulations, we find that a rejection factor of $10^{-13}$ for photons with an energy of 2.8\,GeV to 4\,GeV can be achieved in LDMX. This enables the experiment to reach outstanding sensitivity, allowing to probe several thermal targets in a mass range up to about 100\,MeV with a pilot run of $4\times10^{14}$ electrons on target at a beam energy of 4\,GeV. Later runs at higher energy and with higher statistics are expected to probe a mass range up to several hundred MeV, and to decisively explore a broad range of thermal relic dark matter scenarios.

%% file: appendix/ECalVars.tex
\subsection{Global Variables}

In this Appendix, we show some of variables that provide the highest discrimination power in the BDT.

\begin{figure}[tbh]
\centering
\includegraphics[width=0.48\textwidth]{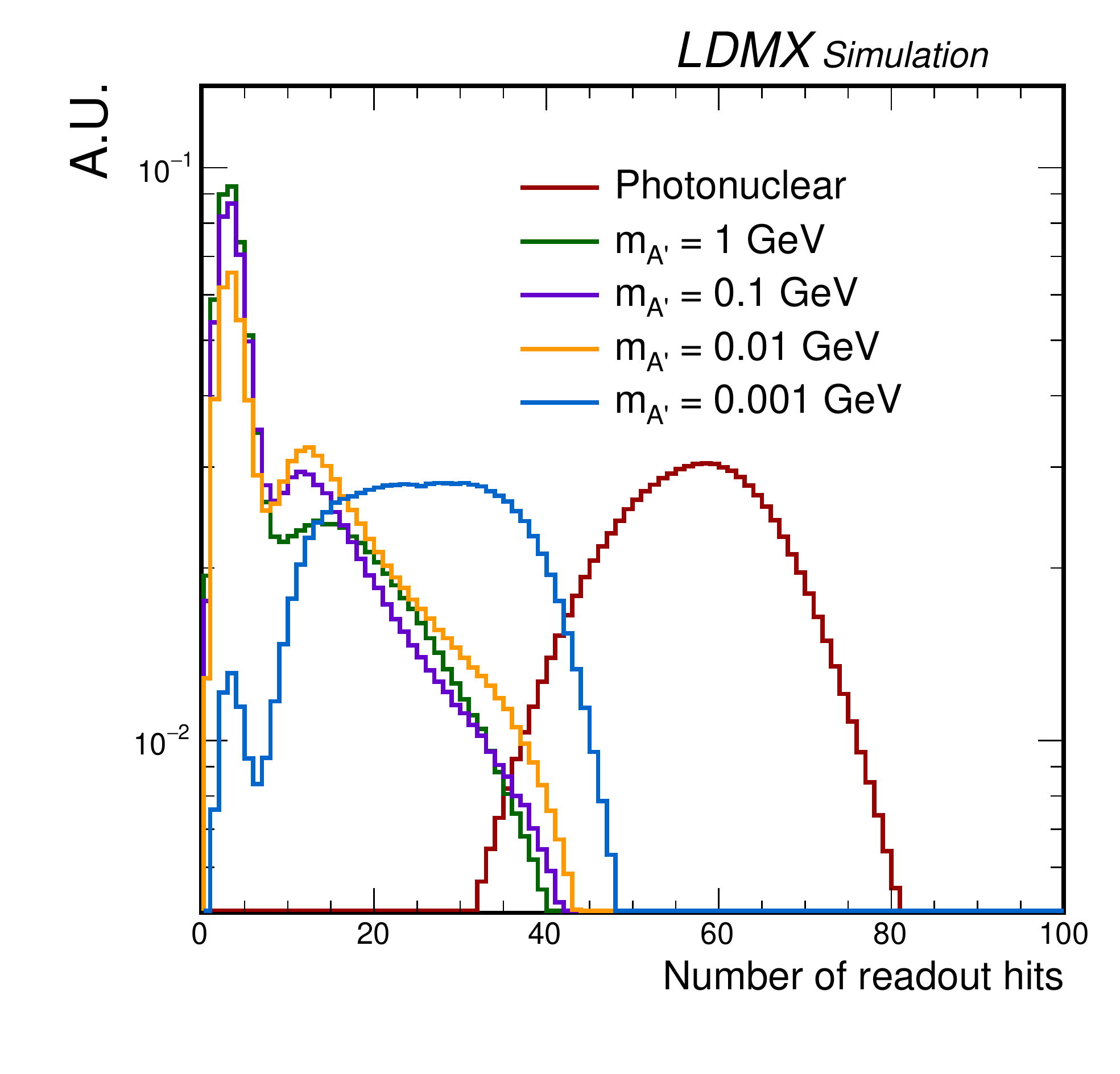}
\includegraphics[width=0.48\textwidth]{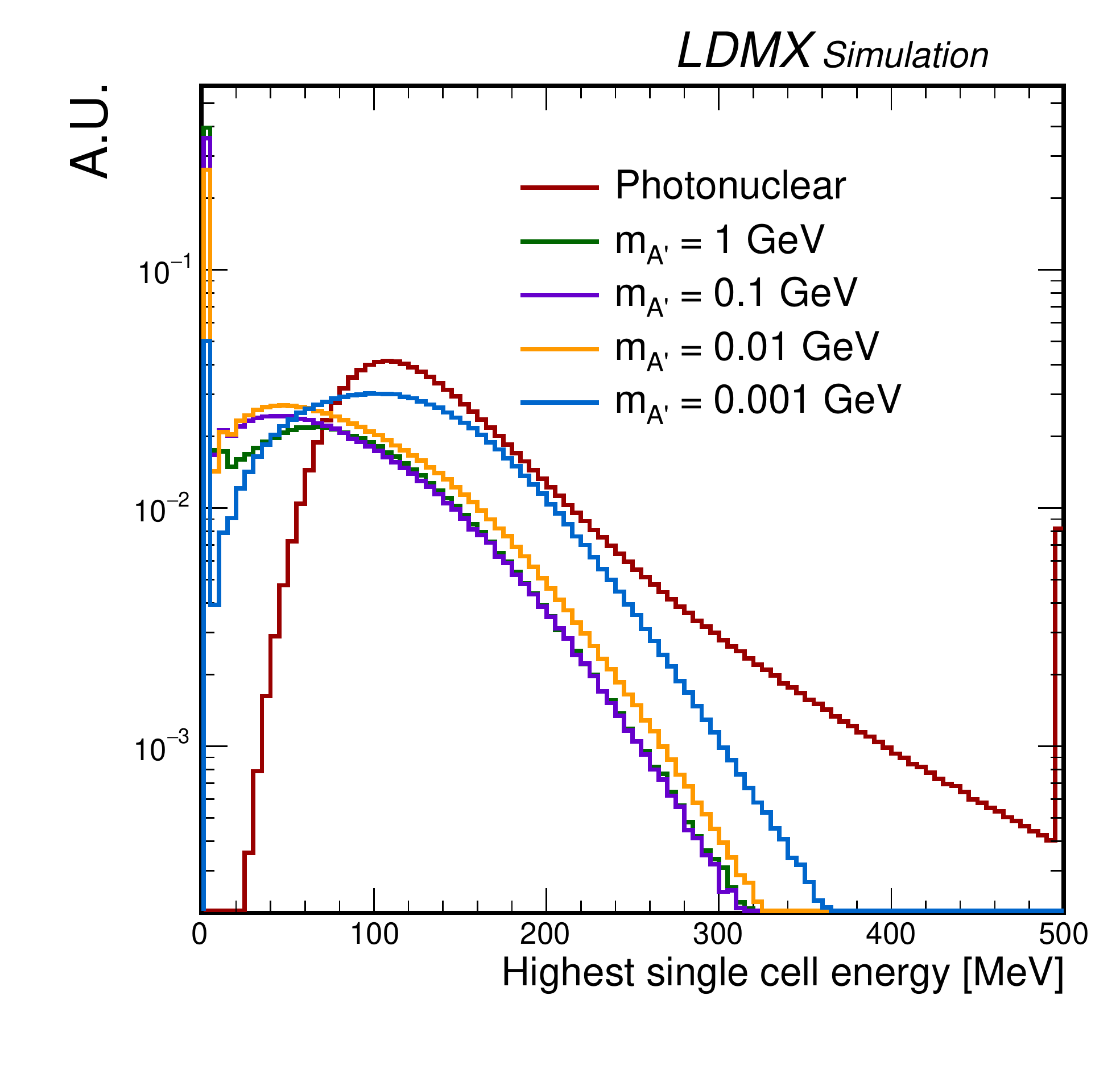}
\begin{flushleft}
\caption{\label{fig:ecalvars_app} Distributions of quantities related to the energy deposited in the \ecal for photo-nuclear and signal processes in events with a total reconstructed energy in the \ecal of less than 1.5\,GeV. Left: Number of hits above the readout threshold. Right: Highest energy in a single cell. The peak close to 0 for the signal is due to events in which the electron is deflected by a wide angle and does not enter the \ecal, resulting in very little energy being observed. All distributions are normalized to unit area. }
\end{flushleft}
\end{figure}

\begin{figure}[tb]
\centering
\includegraphics[width=0.4\textwidth]{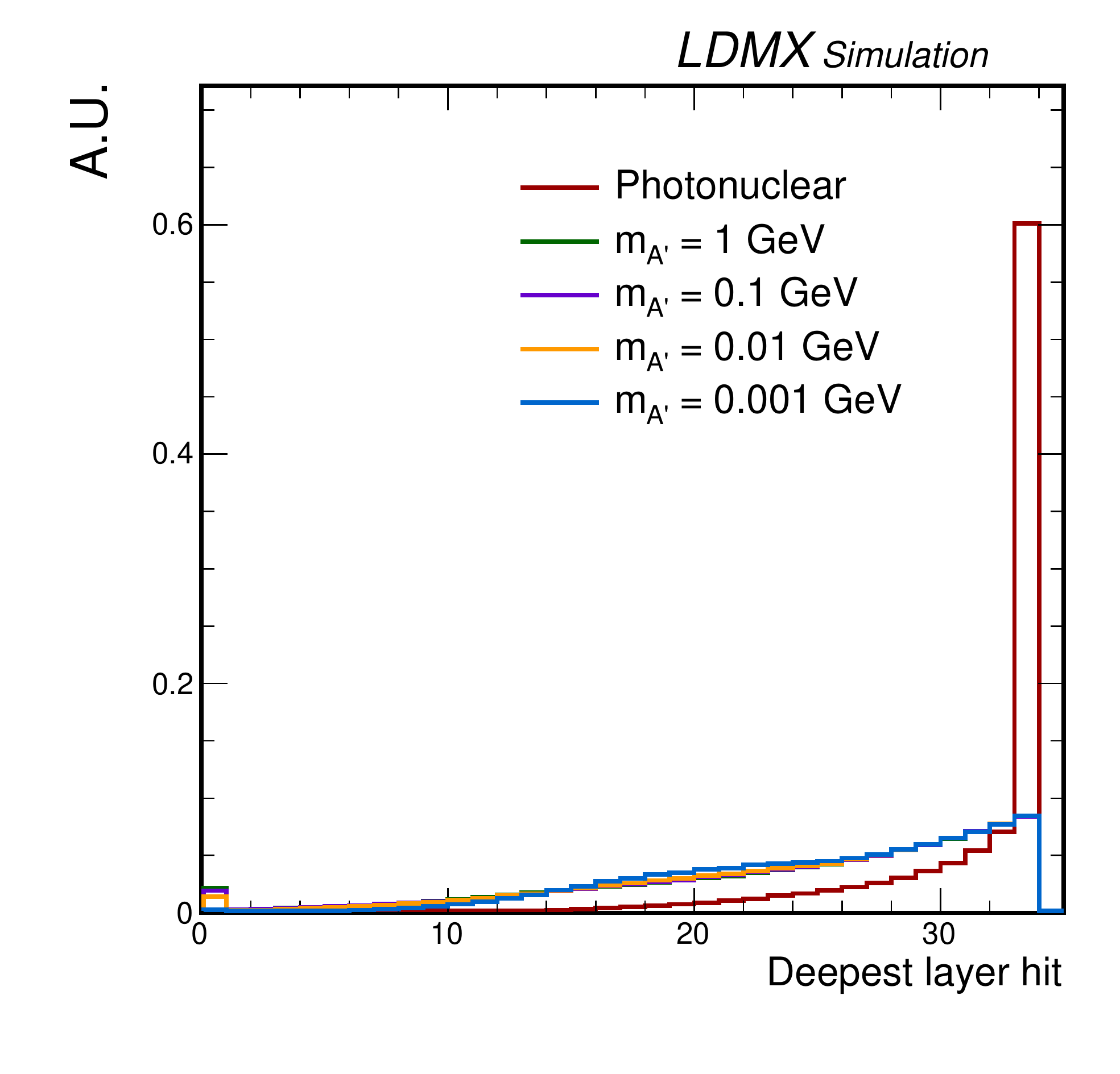} 
\includegraphics[width=0.4\textwidth]{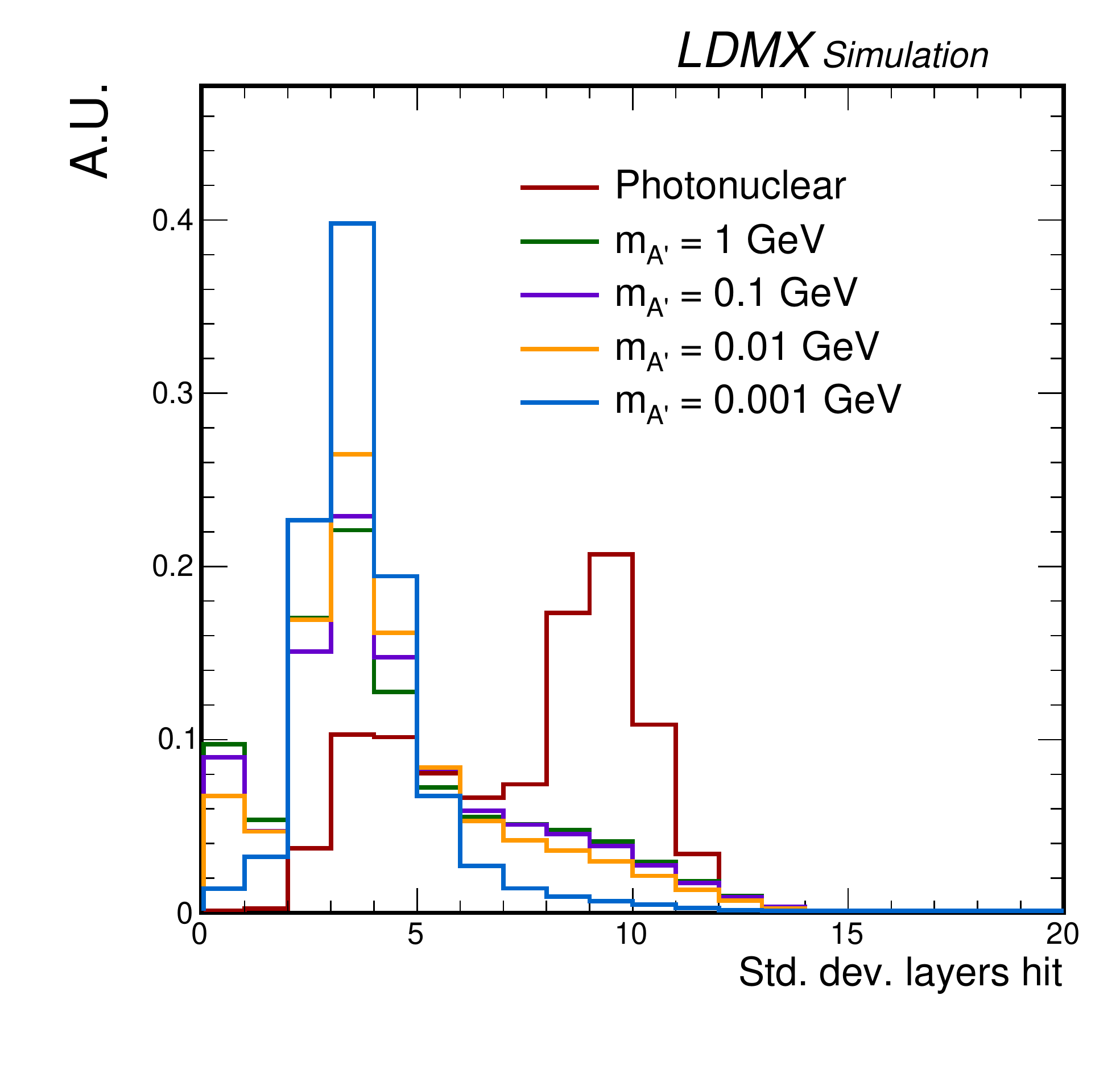}\\
\includegraphics[width=0.4\textwidth]{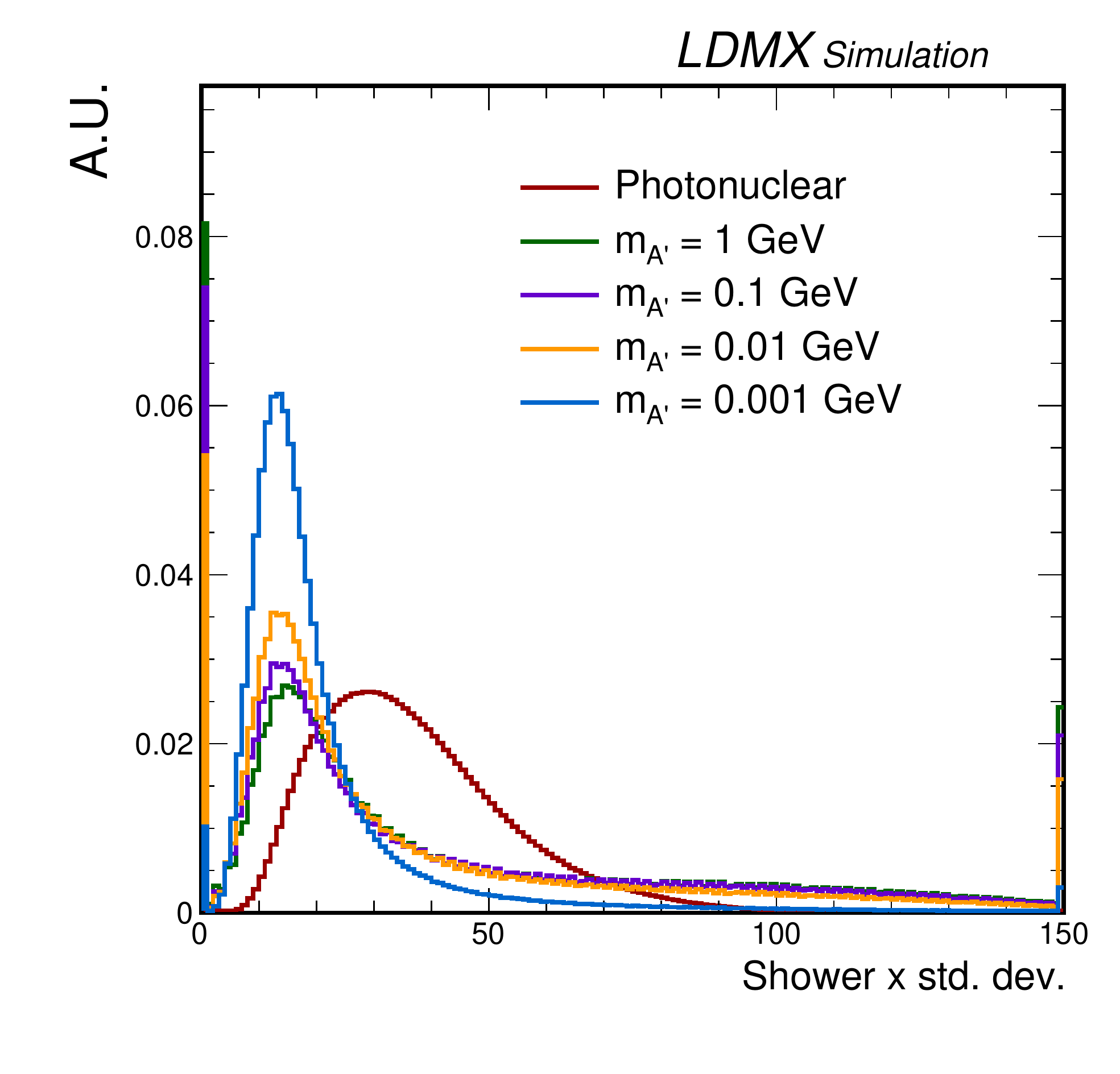}
\includegraphics[width=0.4\textwidth]{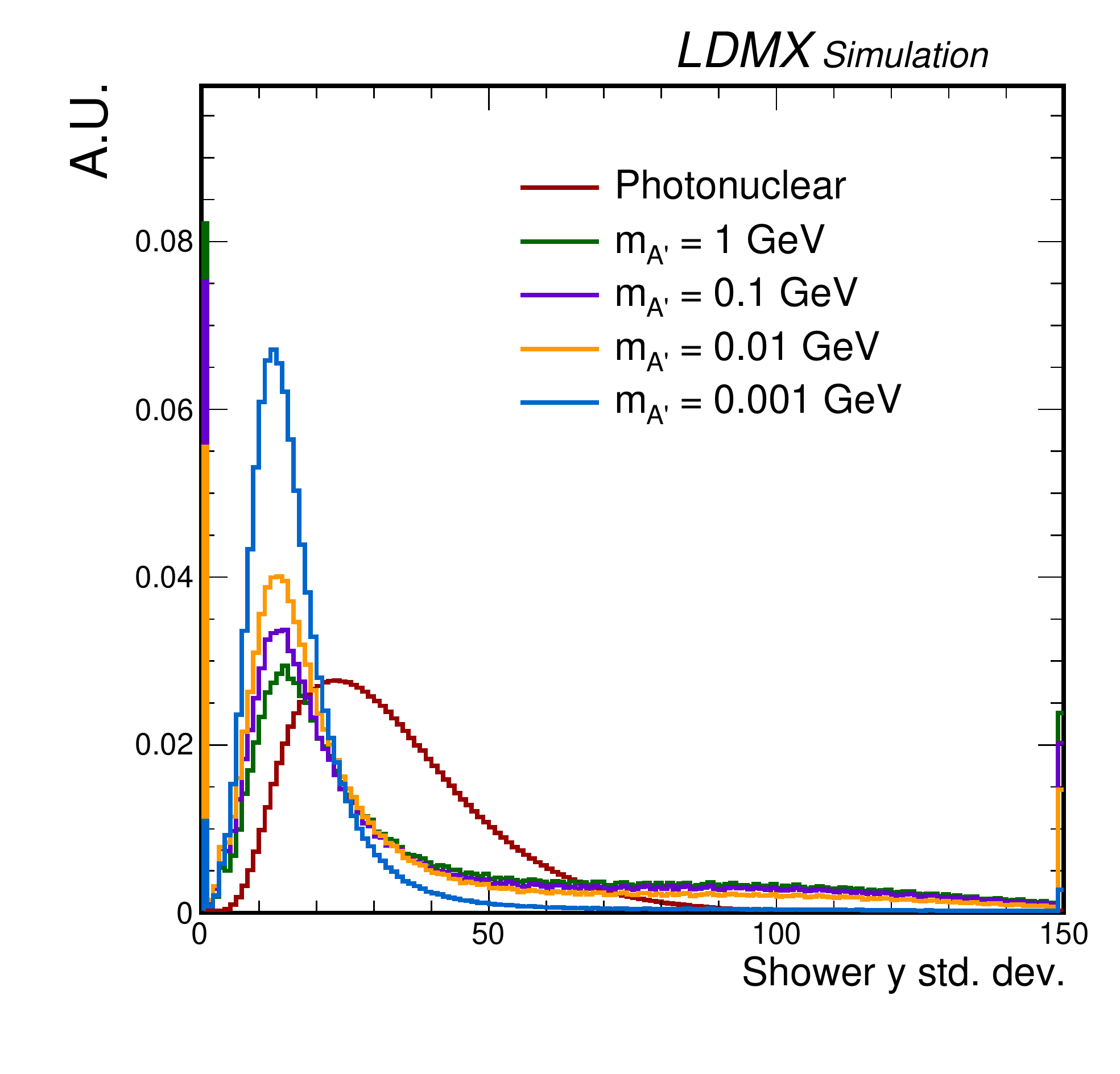}\\
\includegraphics[width=0.4\textwidth]{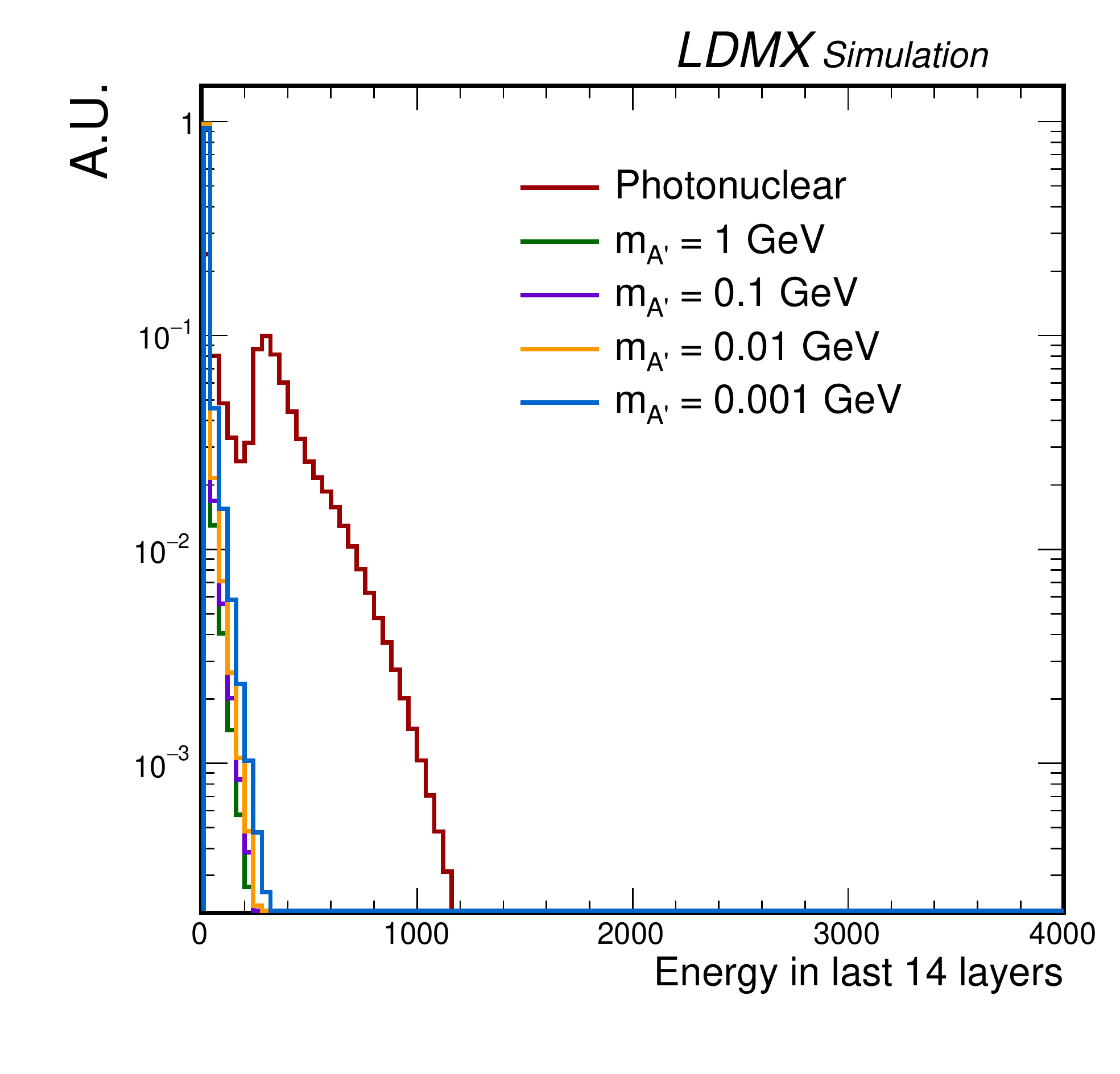}\\
\begin{flushleft}
\caption{\label{fig:bdtTransLong_app} Distributions of quantities related to the longitudinal and transverse shower profiles in the \ecal for photo-nuclear and signal processes in events with a total reconstructed energy in the \ecal of less than 1.5\,GeV. Top: Deepest layer in which a hit is recorded (left) and  standard deviation of the energy-weighted layer numbers (right). Middle: Energy-weighted standard deviation of x-positions of all hits (left), energy-weighted standard deviation of y-positions of all hits (right). The peaks close to 0 for the signal are due to events in which the electron is deflected by a wide angle and does not enter the \ecal, resulting in very few hits being observed. Bottom: Energy in the last 14 layers of the \ecal. All distributions are normalized to unit area.}
\end{flushleft}
\end{figure}

\FloatBarrier
\subsection{Containment Regions}

Tracking information is used to determine the projected trajectory of the recoil electron through the \ecal{}. Under the assumption that the electron emits a hard bremsstrahlung photon in the target, the recoil electron track and kinematics are also used to infer the trajectory of the photon. For each \ecal{} layer, concentric circles centered on these projected electron and photon trajectories are defined, with radii corresponding to integer multiples (up to a maximum of 5) of the 68\% shower containment radius determined for that layer, in order to determine the areas in which electromagnetic shower energy is expected to be reconstructed. 

Five sets of concentric containment regions are thus obtained for each of the trajectories by associating the inner circle and the four successive annuli, defined by the different radii, in each layer. A complementary set of ``outside" regions is defined by combining the areas of the \ecal{} outside the outer radius of each individual ring defined by a given multiple of the 68\% shower containment radius over all the layers. The energy reconstructed from hits in each of these regions is summed, and the multiplicity and energy-weighted standard deviations of the x- and y-positions of hits in each of the outside regions provide additional observables. 

\begin{figure}[tb]
\centering
\includegraphics[width=0.4\textwidth]{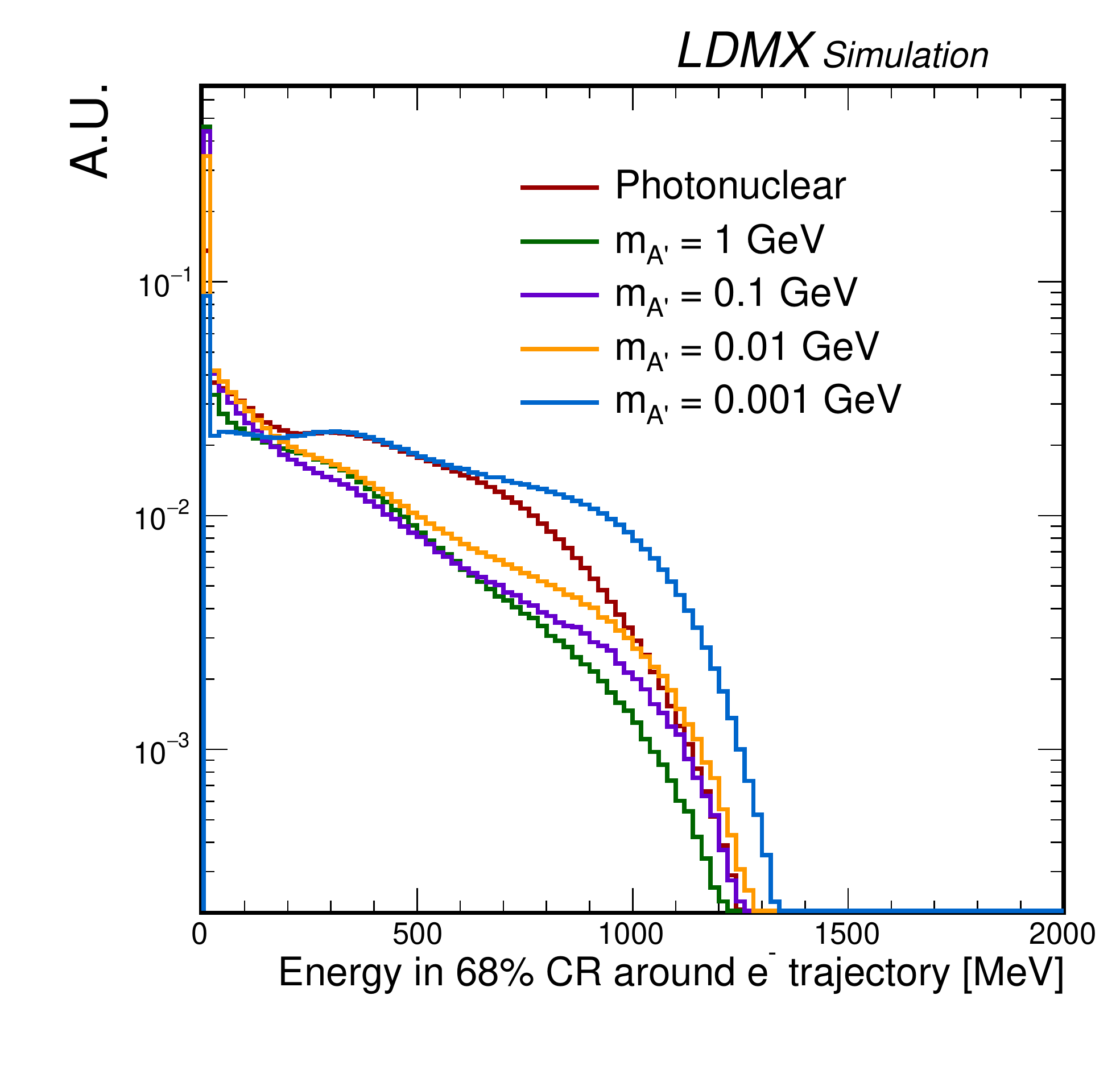}
\includegraphics[width=0.4\textwidth]{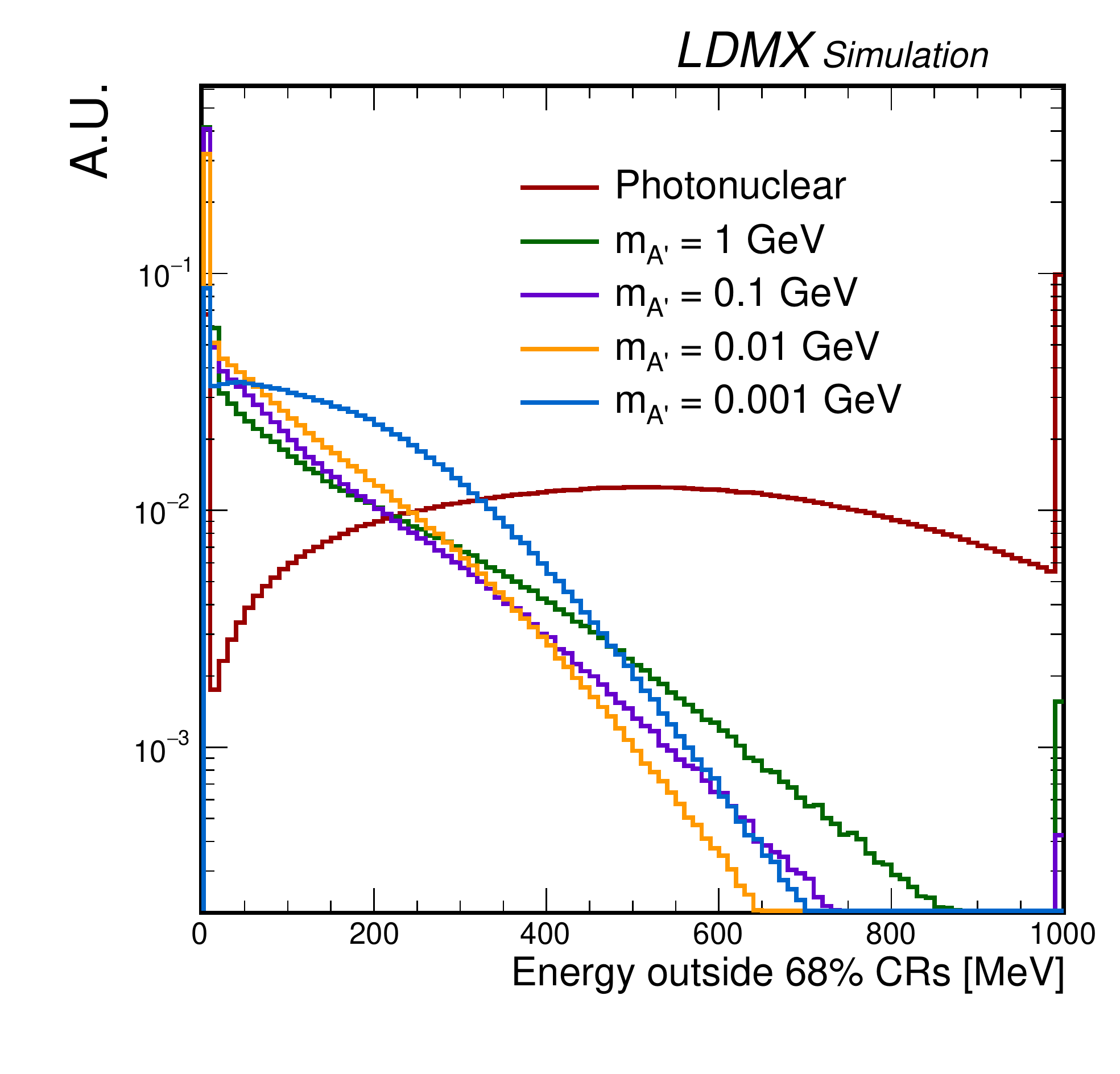}\\
\includegraphics[width=0.4\textwidth]{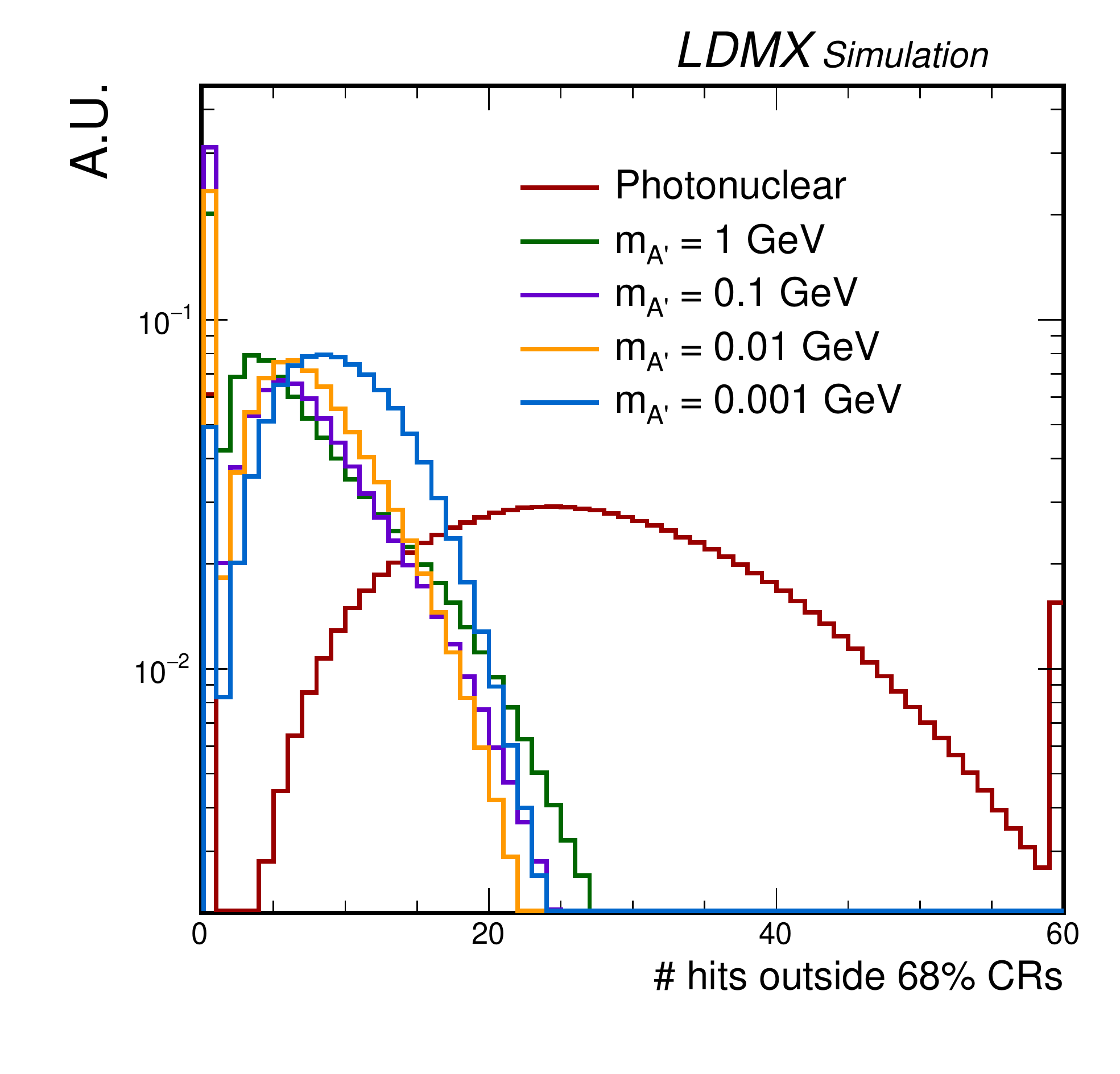}
\includegraphics[width=0.4\textwidth]{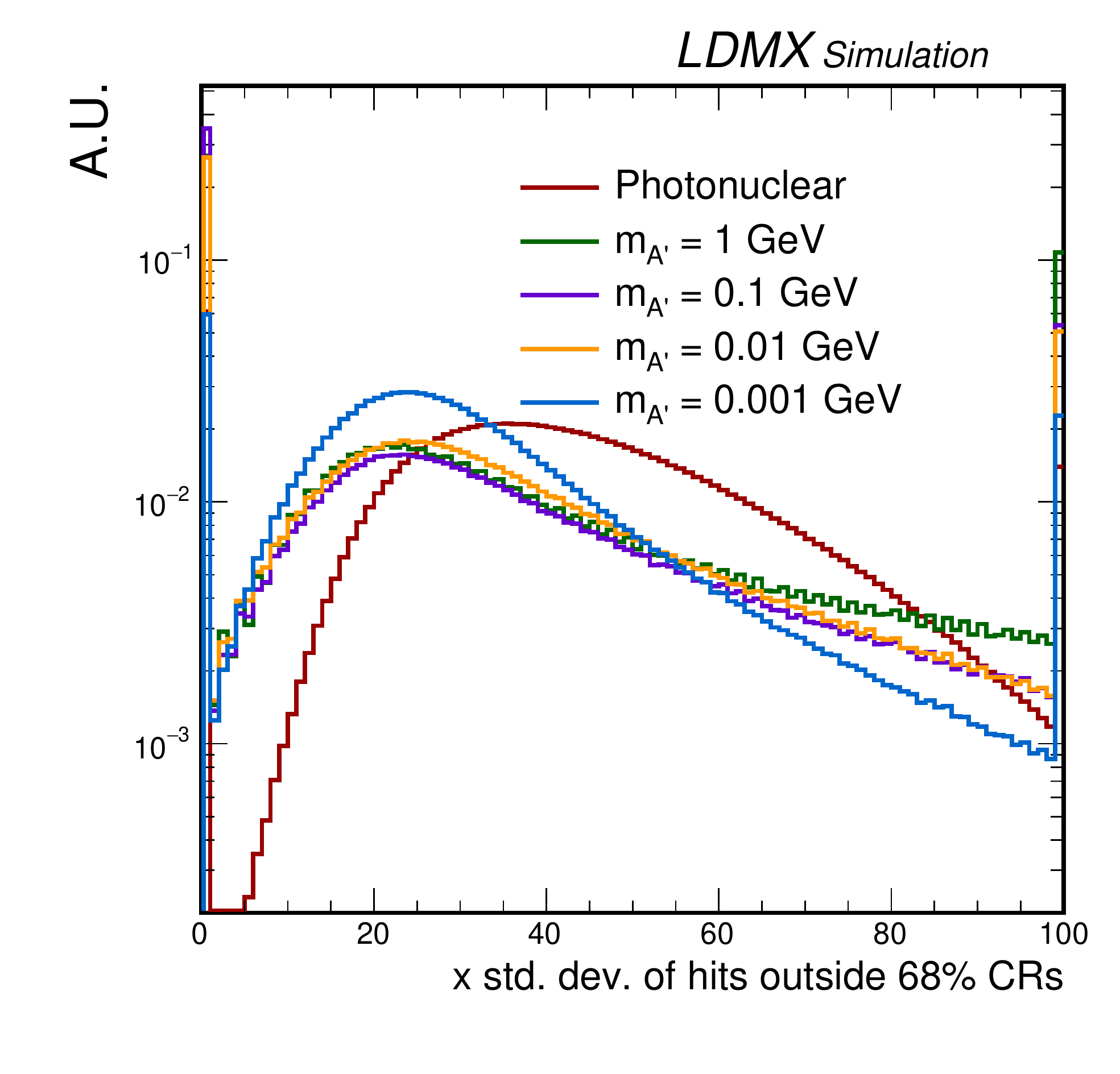}\\
\includegraphics[width=0.4\textwidth]{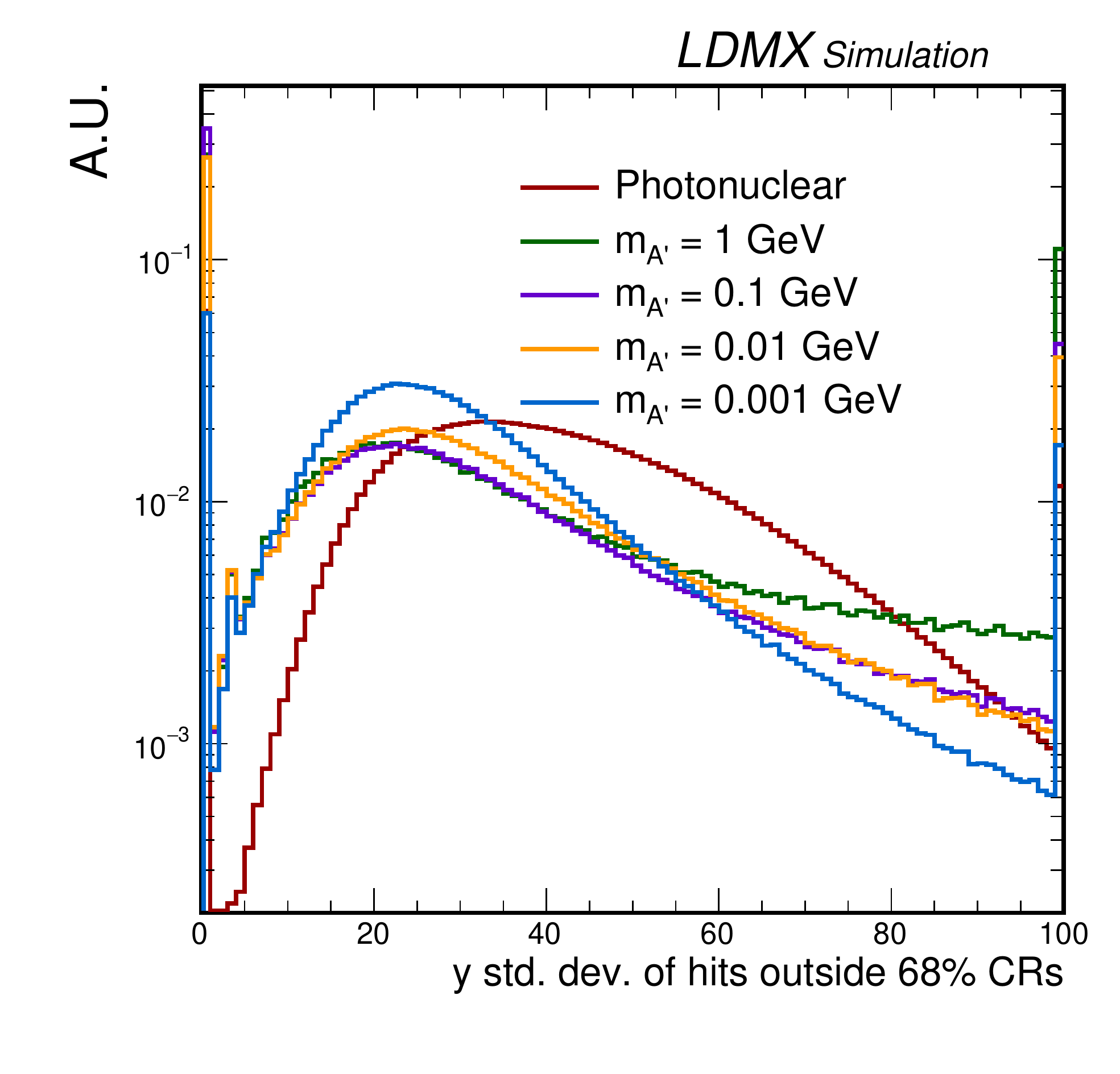}
\begin{flushleft}
\caption{\label{fig:containmentVars_app} Top: Energy reconstructed in containment regions (CRs) defined by 68\% containment radii around electron trajectory (left); energy reconstructed in region outside inner 68\% electron and photon CRs (right). Middle: Number of hits in region outside inner 68\% electron and photon CRs (left), energy-weighted standard deviation of x-positions of hits in region outside inner 68\% electron and photon CRs (right). Bottom: Energy-weighted standard deviation of y-positions of hits in region outside inner 68\% electron and photon CRs. All distributions are normalized to unit area.}
\end{flushleft}
\end{figure}

Some examples are shown in Fig.~\ref{fig:containmentVars_app}.

\FloatBarrier